\documentclass{aa}  
\usepackage{graphicx}

\usepackage{txfonts}
\usepackage{array}
\usepackage{amssymb}
\usepackage{natbib}
\usepackage{lscape}
\usepackage{rotating}
\usepackage{longtable}
\newcommand{\xmm}{{\sc{XMM}}\emph{-{\it Newton}}}
\newcommand{\ch}{{\it Chandra}}
\newcommand{\kms}{km\,s$^{-1}$}
\begin{document}

\title{B stars seen at high resolution by \xmm \thanks{Based on observations collected with the ESA science mission \xmm , an ESA science mission with instruments and contributions directly funded by ESA member states and the USA (NASA).}}

\author{Constantin Cazorla
\and Ya\"el~Naz\'e\thanks{F.R.S.-FNRS Research Associate.}
}

\institute{Groupe d'Astrophysique des Hautes Energies, STAR, Universit\'e de Li\`ege, Quartier Agora (B5c, Institut d'Astrophysique et de G\'eophysique), All\'ee du 6 Ao\^ut 19c, B-4000 Sart Tilman, Li\`ege, Belgium\\
\email{naze@astro.ulg.ac.be}
}

\authorrunning{Cazorla \& Naz\'e}
\titlerunning{B stars in X-rays }
\abstract{We report on the properties of 11 early B stars observed with gratings on board \xmm\ and \ch, thereby doubling the number of B stars analysed at high resolution. The spectra typically appear soft, with temperatures of 0.2--0.6\,keV, and moderately bright ($\log[L_{\rm X}/L_{\rm BOL}]\sim -7$) with lower values for later type stars. In line with previous studies, we also find an absence of circumstellar absorption, negligible line broadening, no line shift, and formation radii in the range 2 -- 7 R$_{\star}$. From the X-ray brightnesses, we derived the hot mass-loss rate for each of our targets and compared these values to predictions or values derived in the optical domain: in some cases, the hot fraction of the wind can be non-negligible. The derived X-ray abundances were compared to values obtained from the optical data, with a fair agreement found between them. Finally, half of the sample presents temporal variations, either in the long-term, short-term, or both. In particular, HD\,44743 is found to be the second example of an X-ray pulsator, and we detect a flare-like activity in the binary HD\,79351, which also displays a high-energy tail and one of the brightest X-ray emissions in the sample.}
\maketitle

\section{Introduction}
The advent of the first high-resolution X-ray spectrographs on board \xmm\ and \ch\ profoundly modified our understanding of high-energy phenomena. Indeed, high-resolution spectra provide a wealth of detailed information. For massive stars, the X-ray emission comes from optically thin, hot plasma; hence lines dominate the X-ray spectra. Their relative strengths closely constrain the plasma temperature and composition. The ratios between forbidden and intercombination lines of He-like ions further pinpoint where X-rays arise \citep{por01}. Finally, their line profiles provide unique information on the stellar wind, its opacity, and its velocity field \citep{mac91,owo01}. 

Analyses of the high-resolution spectra of O stars have revealed winds to be less opaque than initially thought and have helped constraining the properties of high-energy interactions (e.g. colliding winds in binaries and magnetically confined winds in strongly magnetic objects). However, many fewer B stars were observed at high resolution. To further advance the understanding of X-rays associated with early B stars, we present in this paper the X-ray observations of 11 additional targets, thereby doubling the number of such stars observed at high resolution. Section 2 presents the selection of targets, their properties, and their observations. Sections 3 and 4 report our analyses of spectra and light curves, respectively, while Sect. 5 summarises and concludes this paper.

\section{Target selection and observations} 
\label{sect2}
Performing a spectroscopic X-ray survey of early-type B stars at high resolution can only be carried out if the X-ray emission is sufficiently bright; this is the case even with \xmm,\  which is the most sensitive facility currently available. A target selection was therefore performed with the {\it ROSAT} All-Sky Survey (RASS): objects with RASS count rates larger than 0.1\,cts\,s$^{-1}$ yield usable RGS spectra within relatively short exposure times; still, for the great majority
of the observations, these exposure times are greater than 20\,ks. There are 20 such B stars in the catalogue of \citet{ber96}, but one (HD\,152234) actually is an O+O binary \citep[for an analysis of its \xmm\ low-resolution data, see][]{san06} and the high-resolution spectra of 8 stars have been previously analysed.\ These include the strongly magnetic stars HD\,205021 \citep[$\beta$\,Cep, B1III+B6--8+A2.5V;][]{fav09} and HD\,149438 \citep[$\tau$\,Sco, B0.2V;][]{coh03,mew03,wal07,zhe07}, as well as HD\,122451 \citep[$\beta$\,Cen, B1III;][]{raa05}, HD\,116658 \citep[Spica, B1III-IV;][]{zhe07,mil07}, HD\,111123 \citep[$\beta$\,Cru, B0.5III+B2V+PMS;][]{zhe07,wal07,coh08}, HD\,93030 \citep[$\theta$\,Car, B0.2V;][]{naz08}, HD\,37128 \citep[$\epsilon$\,Ori, B0I;][]{wal07,zhe07}, and the peculiar Be star HD\,5394 ($\gamma$\,Cas; \citealt{smi04,lop10,smi12}). This leaves 11 stars, which we analyse here using observations taken by \xmm\ or {\it Chandra} (see Table \ref{journal} for the observing log). This table also presents the main properties of the targets. Amongst these targets, four are known binaries (HD\,63922, HD\,79351, HD\,144217, and HD\,158926; see Table \ref{journal}) while two are known $\beta$\,Cephei pulsators (HD\,44743 and HD\,158926; see Sect. 4 for more details). Six targets were also searched for the presence of magnetic fields.\ HD\,36512 \citep{bag06,gru17}, HD\,36960 \citep{bag06,byc09}, HD\,79351 \citep{hub07}, and HD158926 \citep{byc09} seem non-magnetic while HD\,44743 and HD\,52089 may present very weak fields \citep[$<$100\,G;][]{fos15,nei17}; no strongly magnetic star is thus known in our sample. Finally, there is no Be star amongst them, hence none can belong to the class of peculiar, X-ray bright $\gamma$-Cas objects. We thus refrain from comparing our targets to such objects in the following, especially since their X-ray emission has a completely different origin than for normal B stars \citep[for a review, see][]{smi16}. We make comparisons, however, with the published analyses of the 7 other B stars.

\begin{sidewaystable*}
\centering
\caption{Journal of the observations and physical properties of the sample stars.}
\label{journal}
\setlength{\tabcolsep}{2pt}
\begin{tabular}{lcccccccccccc}
\hline\hline
Target name                             & Sp. type      & $d$   & log($L_{\rm BOL}/L_{\odot}$)        &$R_{\star}$    &$T_{\rm eff}$  &$N^{ism}_{\rm H}$         & ObsID (Rev)                   & Obs. mode     & PI            & Start date                      & Flare?        & Duration \\
                                                &                       & (pc)    &                                               &($R_{\odot}$)  &(K)                    & ($10^{22}$\,cm$^{-2}$)  &                                       &                        &               &                                       &         & (ks)\\
\hline
HD\,34816 ($\lambda$\,Lep)      & B0.5 V     [1] *      & 261           & 4.22                                    &4.7            &30400          & 1.66e-2                                 & 0690200601 (2415)     & ff+thick                 & Naz\'e        & 2013-02-15@07:41:26   & N             & 26.5 \\
HD\,35468 ($\gamma$\,Ori)       & B2 IV--III [2] *      & 77            & 3.87                                    &6.4            &21250          & 1.82e-3                                 & 0690680501 (2342)     & ff+thick                 & Waldron& 2012-09-22@19:27:17  & Y             & 42.0 \\
HD\,36512 ($\upsilon$\,Ori)     & B0 V       [1] *      & 877           & 5.21                                    &12.1           &33400          & 1.95e-2                                 & 0690200201 (2326)     & ff+thick                 & Naz\'e        & 2012-08-22@02:46:15   & Y             & 20.2 \\
HD\,36960                               & B0.7 V     [1] *      & 495           & 4.58                                    &7.8            &29000          & 2.24e-2                                 & 0690200501 (2433)     & ff+thick                 & Naz\'e        & 2013-03-23@08:57:52   & Y             & 28.2 \\
HD\,38771 ($\kappa$\,Ori)       & B0.5 Ia    [3]                & 198           & 4.72                                    &12.5           &24800          & 3.39e-2                                 & 9940                                  & ACIS-S          & Waldron& 2008-12-26@23:50:00  &               & 234.2\\
                                                &                               &         &                                               &                       &                       &                                         & 10846                         & +HETG,          &               & 2008-12-30@02:53:17\\
                                                &                               &                 &                                               &                       &                       &                                         & 9939                                  & faint                   &               & 2009-01-03@03:00:24\\
                                                &                               &         &                                               &                       &                       &                                         & 10839                                 &                         &               & 2009-01-05@18:25:49\\
HD\,44743 ($\beta$\,CMa)        & B1 II--III [4]                & 151           & 4.41                                    &8.8                    &24700          & 2.00e-4                                 & 0503500101 (1509)     & ff+thick         & Waldron& 2008-03-06@15:33:36  & Y             & 10.6 \\
                                                &                               &         &                                               &                       &                       &                                         & 0761090101 (2814)     & ff+thick                 &Oskinova& 2015-04-21@06:44:53  & Y             & 78.0 \\
HD\,52089 ($\epsilon$\,CMa)     & B1.5 II    [4]                & 124           & 4.35                                    &9.9                    &22500          & 1.00e-4                                 & 0069750101 (0234)     & {\tiny ff(MOS)/sw(pn)+thick}& Cohen  & 2001-03-19@12:05:51 & Y & 31.2 \\
HD\,63922 (P\,Pup)                      & B0.2 III+? [1,5] *    & 505           & 4.95                                    &10.4           &31200          & 6.71e-2                                 & 0720390601 (2552)     & ff+thick                 & Naz\'e        & 2013-11-15@02:11:45   & N             & 46.5 \\
HD\,79351 (a\,Car)                      & B2.5 V+?   [6,7] * & 137      & 3.60                                    &5.8            &19150          & 6.04e-2                                 & 0690200701 (2393)     & ff+thick         & Naz\'e        & 2013-01-02@00:12:30   & N             & 47.1 \\
HD\,144217 ($\beta^1$\,Sco)     & {\tiny B0.5IV-V+B1.5V [8] *} & 124   & 4.44                            &6.1            &30400          & 1.37e-1                                 & 0690200301 (2425)     & ff+thick               & Naz\'e        & 2013-03-06@16:04:28   & N             & 26.6 \\
HD\,158926 ($\lambda$\,Sco)     & B1.5 IV+ (DA.79 \\
                                                & or PMS)+B [9,10] *& 175       & 4.73                                    &15.3           &22500          & 2.51e-3                                 & 0690200101 (2424)     & ff+thick         & Naz\'e & 2013-03-04@15:37:29  & N             & 17.8 \\
\hline
\end{tabular}
\\
\tablefoot{References for spectral types: [1] \citet{nie13}; [2] \citet{mor08}; [3] \citet{len92}; [4] \citet{fos15}; [5] \citet{nie14}; [6] \citet{sle82}; [7]  \citet{bus60}; [8] \citet{hol97}; [9] \citet{hol13}; [10] \citet{uyt04a}. Distances, based on Hipparcos, come from \citet{van07} and interstellar absorbing columns $N^{ism}_{\rm H}$ were taken from Table 2 (the most recent values) of \citet{gud12}, except for HD\,36960 (for which there is no available value in \citealt{gud12} so the adopted value is taken from \citealt{dip94}) and HD\,63922, for which only the colour excess is known \citep[$E(B-V)=0.11$;][]{kat13}, which was converted into an absorbing column using the relation of \citet{gud12}. Bolometric luminosities (for the primary star if binary) were calculated using their Simbad V magnitudes, Hipparcos distances, reddenings from \citet[][except for HD\,63922, see above]{gud12}, and bolometric corrections of \citet{nie13} for non-supergiant stars (identified by *) and the scale of \citet{cox02} for HD\,38771. Effective temperatures $T_{\rm eff}$ were also taken from the latter two references with the addition of \citet{fos15} for HD\,44743 and HD\,52089 (these authors also provided $L_{\rm BOL}$ values for those stars, which are quoted in a previous column). Radii were calculated from $L_{\rm BOL}=4\,\pi\,R_{\star}^{2}\,\sigma\,T_{\rm eff}^{4}$. For the EPIC cameras, ``ff'' and ``sw'' correspond to full frame and small window modes, respectively; the thick filter was always needed to reject optical/UV light in view of the (optical/UV) brightness of the targets. The durations in the last column correspond to exposure times after flare filtering for EPIC-pn or ACIS-S (0th order). The $\kappa$\,Ori data, taken within 10 days, were combined, so that only the total duration is provided.}
\end{sidewaystable*}

\subsection{\xmm}
The \xmm\ data were reduced with {the Science Analysis Software (SAS)} v16.0.0 using calibration files available in Spring 2016 and following the recommendations of the \xmm\ team\footnote{SAS threads, see \\ http://xmm.esac.esa.int/sas/current/documentation/threads/ }. 

After the initial pipeline processing, the European Photon Imaging Camera (EPIC) observations were filtered to keep only the best-quality data ({\sc{pattern}} 0--12 for MOS and 0--4 for pn). Light curves for events beyond 10\,keV were calculated for the full cameras and, whenever background flares were detected, the corresponding time intervals were discarded; we used thresholds of 0.2\,cts\,s$^{-1}$ for MOS, 0.5\,cts\,s$^{-1}$ for pn in full-frame mode and 0.04\,cts\,s$^{-1}$ for pn in small window mode. We extracted EPIC spectra using the task {\it{especget}} in circular regions of 50\arcsec\ radius for MOS and 35\arcsec\ for pn (to avoid CCD gaps) centred on the Simbad positions of the targets except for HD\,36960 and HD\,144217, for which the radii were reduced to 25\arcsec\ and 7.5\arcsec\, respectively, owing to the presence of neighbouring sources. Dedicated {Ancillary Response File (ARF)} and {Redistribution Matrix File (RMF)} response matrices, which are used to calibrate the flux and energy axes, respectively, were also calculated by this task. We grouped EPIC spectra with {\it{specgroup}} to obtain an oversampling factor of five and to ensure that a minimum signal-to-noise ratio of 3 (i.e. a minimum of 10 counts) was reached in each spectral bin of the background-corrected spectra; unreliable bins below 0.25\,keV were discarded. We extracted EPIC light curves in the same regions as the spectra, for time bins of 100\,s and 1\,ks,  and in the 0.3--10.0\,keV energy band. These light curves were further processed by the task {\it epiclccorr}, which corrects for the loss of photons due to  vignetting, off-axis angle, or other problems such as bad pixels. In addition, to avoid very large errors, we discarded bins displaying effective exposure time lower than 50\% of the time bin length. 

{The Reflection Grating Spectrometer (RGS)} data were also locally processed using the initial SAS pipeline. As for EPIC data, a flare filtering was then applied (using a threshold of 0.1\,cts\,s$^{-1}$). For HD\,36960 and HD\,144217, close companions exist at a distance of 0.14--0.62' and 0.2', respectively, perpendicular to the dispersion axis. The extraction region was thus reduced to 45\% and 50\% of the PSF radius, respectively, to avoid contamination. Furthermore, the background was extracted after excluding regions within 98\% of the PSF size from the source and neighbours. Dedicated response files were calculated for both orders and both RGS instruments and were subsequently attached to the source spectra for analysis. The two RGS datasets of HD\,44743 were combined using {\it rgscombine}. For individual line analyses, no grouping and no background were considered (the local background around the lines was simply fitted using a flat power law). For global fits, background spectra were considered and a grouping was performed to reach at least 10\,cts per bin and to get emissions of each He-like ($fir$) triplet in a single bin. The latter step is needed because global spectral fitting does not take into account the possible depopulation of the upper level of the $f$-line in triplets in favour of the upper levels of the  $i$-lines. 

\subsection{\ch}
The \ch\ data of HD\,38771 were reprocessed locally using CIAO 4.9 and CALDB 4.7.3. Since all \ch\ data were taken within 10 days, orders +1 and --1 were added and the four exposures combined using {\it combine\_grating\_spectra} to get the final HEG and MEG spectra. For each exposure, 0th order spectra of the source were also extracted in a circle of 2.5\arcsec\ radius around its Simbad position, while the associated background spectra were extracted in the surrounding annulus with radii 2.5 and 7.4\arcsec. Dedicated response matrices were calculated using the task {\it specextract}, and the individual spectra and matrices were then combined using the task {\it combine\_spectra} to get a single spectrum for HD\,38771. The resulting HEG, MEG, and 0th order spectra were grouped in the same way as the \xmm\ spectra. For each exposure, light curves were extracted considering the same regions as the spectra, for time bins of 100\,s and 1\,ks, and in the 0.3--10.0\,keV energy band. 

\section{Spectra}
We performed two types of fitting: individual analyses of the lines and global analyses. All these analyses were performed within Xspec v 12.9.0i.

\begin{figure*}
\includegraphics[width=9.5cm]{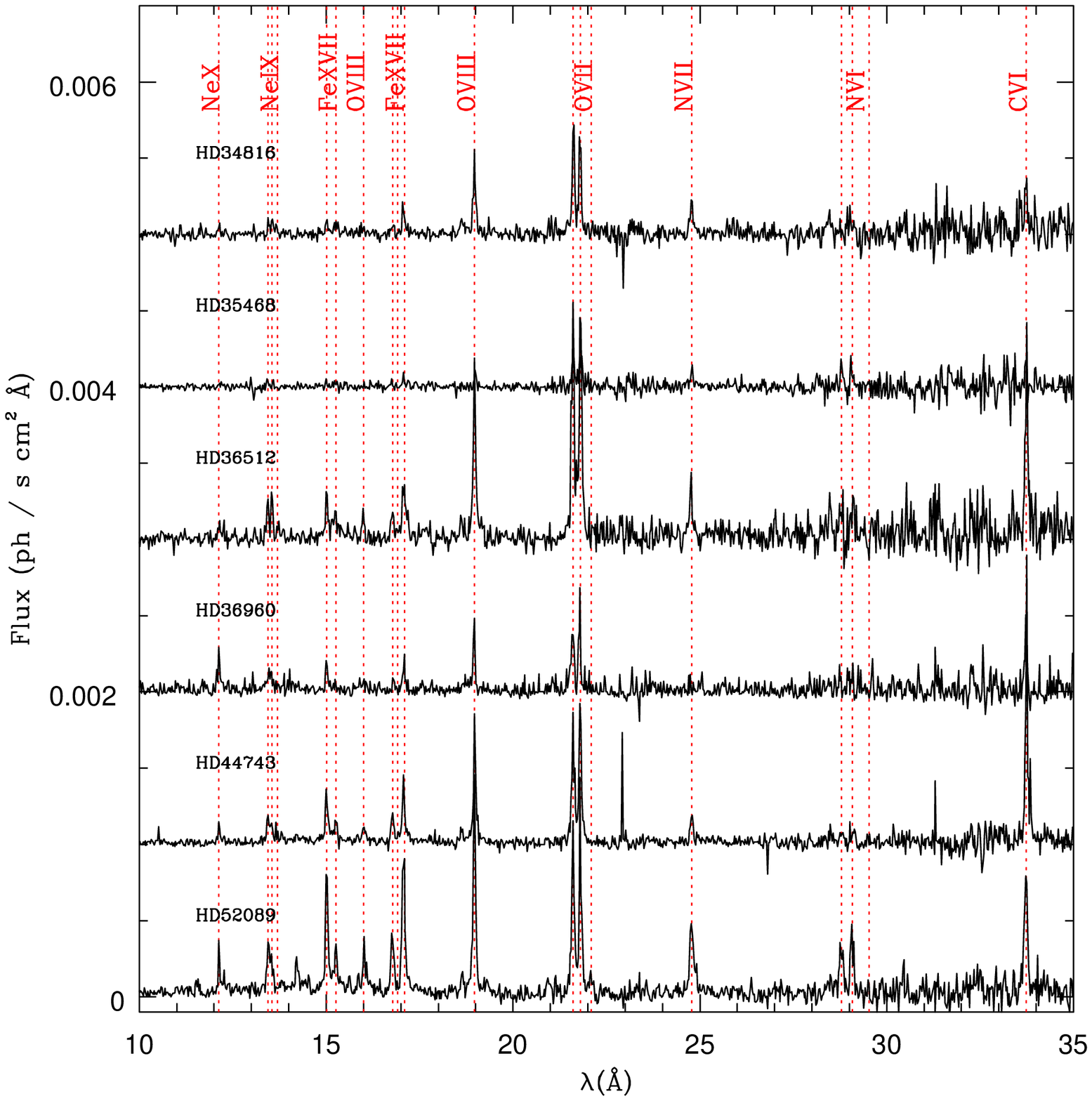}
\includegraphics[width=9.5cm]{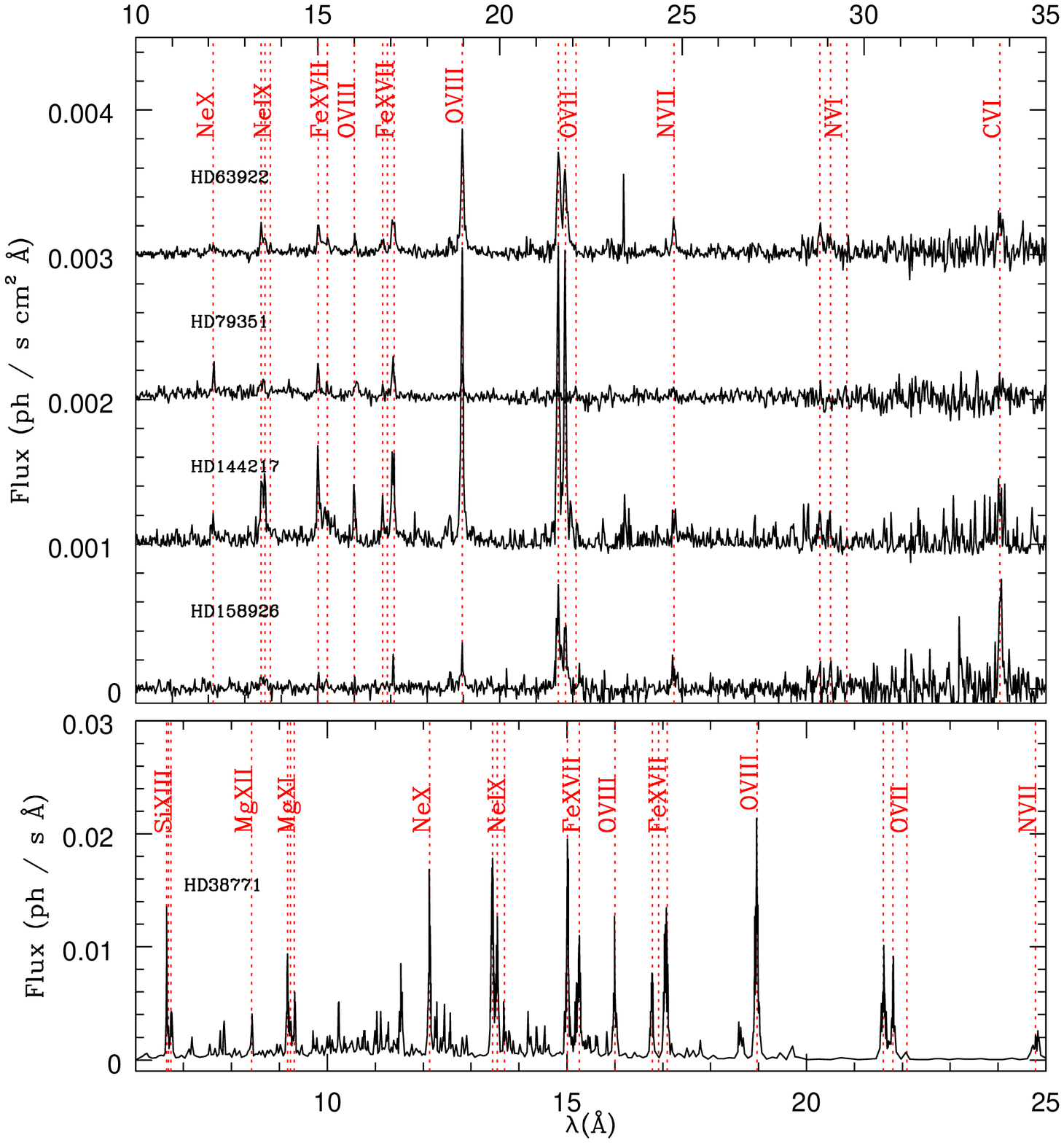}
\caption{RGS and MEG spectra of our targets with the main lines identified and their position indicated by the dotted lines.}
\label{highres}
\end{figure*}

\subsection{Line fits}
\label{sec3_}
Figure \ref{highres} shows the high-resolution spectra of our targets. Lines from H-like and He-like ions of C, N, O, Ne, Mg, and Si, along with lines from ionised Fe, are readily detected, demonstrating the thermal nature of the X-ray emissions. In general, the triplet from He-like ions appear stronger than Ly$\alpha$ lines, as expected for late-type massive stars \citep{wal09}, but we find no clear, systematic dependence with the spectral types of our targets. 

An examination of the line profiles reveals no obvious asymmetry, hence we decided to fit them by simple Gaussian profiles. We relied on Cash statistics and therefore used unbinned spectra without background correction. We only fitted the lines that were sufficiently intense to provide a meaningful fit. In case of doublets (the two components of Lyman lines of H-like ions) or triplets (the $fir$ components of He-like ions), the individual components were forced to share the same velocity and the same width. Furthermore, the flux ratios between the two Lyman components and the flux ratios between the two $i$ lines of Si and Mg\footnote{For N, O, and Ne, the second $i$ component is $>$30 times less intense, hence can be neglected.} were fixed to the theoretical ratios in ATOMDB\footnote{See e.g. http://www.atomdb.org/Webguide/webguide.php}. The derived line properties are listed in Table \ref{linefit} with 1$\sigma$ errors determined using the {\it error} command under Xspec. Whenever the fitted width was reaching a null value, a second fitting was performed with the width fixed to zero, and Table \ref{linefit} provides the results of this second fitting.

\begin{sidewaystable*}
{
\centering
\caption{Results of the individual line fitting by Gaussians. }
\label{linefit}
\begin{tabular}{lccccccccccc}
\hline\hline
& HD\,34816 & HD\,35468 & HD\,36512 & HD\,36960 & HD\,38771 & HD\,44743 & HD\,52089 & HD\,63922 & HD\,79351 & HD\,144217 & HD\,158926 \\
\hline
{\it He-like triplets}\\
\multicolumn{2}{l}{Si\,{\sc xiii}\,$\lambda$6.648,6.685,6.688,6.740\AA } \\
$v$ (\kms)   & &&&& 136$\pm$90 \\
$FWHM$ (\kms)& &&&& 1201$\pm$182 \\
$F_r$        & &&&& 2.72$\pm$0.33 \\
$F_i$        & &&&& 0.62$\pm$0.19 \\
$F_f$        & &&&& 1.29$\pm$0.23 \\
$f/i$        & &&&& 2.08$\pm$0.74 \\
$(f+i)/r$   & &&&& 0.70$\pm$0.14 \\
$T_{fir}$ (keV)     &&&&& 1.30$\pm$0.81 \\ 
\multicolumn{2}{l}{Mg\,{\sc xi}\,$\lambda$9.169,9.228,9.231,9.314\AA } \\
$v$ (\kms)   & &&&& --34$\pm$57 \\
$FWHM$ (\kms)& &&&& 1132$\pm$108 \\
$F_r$        & &&&& 5.45$\pm$0.55 \\
$F_i$        & &&&& 3.00$\pm$0.48 \\
$F_f$        & &&&& 2.87$\pm$0.45 \\
$f/i$        & &&&& 0.96$\pm$0.21 \\
$(f+i)/r$               & &&&& 1.08$\pm$0.16 \\
$T_{fir}$ (keV)         & &&&& 0.25$\pm$0.08 \\
$R_{fir}$ ($R_{\star}$)     & &&&& 2.38$\pm$0.44 \\
\multicolumn{12}{l}{Ne\,{\sc ix}\,$\lambda$13.447,13.553,13.699\AA }\\
$v$ (\kms)   & && --160$\pm$67 && --54$\pm$44 & --95$\pm$110 & 24$\pm$130 & --54$\pm$109 && --37$\pm$86 \\
$FWHM$ (\kms)& && 0 (fixed) && 1310$\pm$71 & 732$\pm$309 & 830$\pm$291 & 369$\pm$441 && 0 (fixed) \\
$F_r$        & && 22.9$\pm$4.5 && 50.1$\pm$3.0 & 17.3$\pm$2.4 & 33.8$\pm$4.6 & 17.5$\pm$2.5 && 38.0$\pm$7.4 \\
$F_i$        & && 26.7$\pm$4.7 && 31.7$\pm$2.9 & 11.2$\pm$2.2 & 19.5$\pm$4.2 & 12.1$\pm$2.4 && 41.5$\pm$8.4 \\
$F_f$        & && 6.53$\pm$3.38 && 7.97$\pm$1.62 & 3.28$\pm$1.47 & 0.46$\pm$2.59 & 5.56$\pm$1.89 && 4.36$\pm$4.56 \\
\multicolumn{12}{l}{O\,{\sc vii}\,$\lambda$21.602,21.804,22.098\AA }\\
$v$ (\kms)   & 1$\pm$146 & 142$\pm$111 & 0$\pm$57 & --254$\pm$98 & --2$\pm$47 & --31$\pm$31 & 0$\pm$60 & 79$\pm$61 && --135$\pm$61 & 0$\pm$108 \\
$FWHM$ (\kms)& 0 (fixed) & 0 (fixed) & 361$\pm$353 & 313$\pm$406 & 1353$\pm$80 & 426$\pm$198 & 0 (fixed) & 1033$\pm$144 && 0 (fixed) & 845$\pm$391\\
$F_r$        & 66.0$\pm$8.3 & 18.7$\pm$4.1 & 161$\pm$17 & 44.1$\pm$9.9 & 300$\pm$27 & 84.5$\pm$6.8 & 107$\pm$8 & 102$\pm$8 && 204$\pm$19 & 82.9$\pm$10.0\\
$F_i$        & 73.1$\pm$8.6 & 28.8$\pm$4.7 & 175$\pm$20 & 53.8$\pm$10.3 & 238$\pm$24 & 107$\pm$7 & 117$\pm$9 & 90.6$\pm$7.9 && 210$\pm$20 & 57.2$\pm$9.9\\
$F_f$        & 2.87$\pm$3.95 & 0 (fixed) & 3.08$\pm$6.03 & 0.20$\pm$4.65 & 25.2$\pm$10.3 & 0.$\pm$0.73 & 6.80$\pm$3.95 & 1.18$\pm$3.13 && 4.59$\pm$6.96 & 8.65$\pm$5.94 \\
$f/i$        & 0.04$\pm$0.05 & 0 (fixed) & 0.02$\pm$0.03 & 0.0$\pm$0.09 & 0.11$\pm$0.04 & 0.000$\pm$0.007 & 0.06$\pm$0.03 & 0.01$\pm$0.03 && 0.02$\pm$0.03 & 0.15$\pm$0.11 \\
$(f+i)/r$               & 1.15$\pm$0.20                         & 1.55$\pm$0.42         & 1.11$\pm$0.17                                 & 1.22$\pm$0.38         & 0.88$\pm$0.12                           & 1.26$\pm$0.13         & 1.16$\pm$0.13                                 & 0.90$\pm$0.11                                 &       & 1.05$\pm$0.14                           & 0.79$\pm$0.17\\
$T_{fir}$ (keV)         &0.09$\pm$0.05  & <0.09         & 0.10$\pm$0.03         & 0.09$\pm$0.08   & 0.22$\pm$0.08         & <0.09 & 0.09$\pm$0.03 & 0.18$\pm$0.06 &       & 0.12$\pm$0.05   & 0.26$\pm$0.18\\
$R_{fir}$ ($R_{\star}$)     &<10.1 &  & <8.3 & <9.8 & 7.4$\pm$2.3 & <2.1 & 4.5$\pm$2.4 & <6.9 &  & <7.6 & 7.3$\pm$6.5 \\
\multicolumn{12}{l}{N\,{\sc vi}\,$\lambda$28.787,29.084,29.535\AA } \\
$v$ (\kms)   & &&&&&& --127$\pm$76 \\
$FWHM$ (\kms)& &&&&&& 685$\pm$182  \\
$F_r$        & &&&&&& 42.8$\pm$5.9\\
$F_i$        & &&&&&& 49.3$\pm$9.2\\
$F_f$        & &&&&&& 0.0$\pm$3.1\\
$f/i$        & &&&&&& 0.00$\pm$0.06\\
$(f+i)/r$               & &&&&&& 1.15$\pm$0.28\\
$T_{fir}$ (keV)         & &&&&&& 0.09$\pm$0.08\\
$R_{fir}$ ($R_{\star}$)    & &&&&&& <3.7\\
\hline
\end{tabular}
}
\end{sidewaystable*}
\setcounter{table}{1}
\begin{sidewaystable*}
{\footnotesize
\centering
\caption{Continued }
\begin{tabular}{lccccccccccc}
\hline\hline
& HD\,34816 & HD\,35468 & HD\,36512 & HD\,36960 & HD\,38771 & HD\,44743 & HD\,52089 & HD\,63922 & HD\,79351 & HD\,144217 & HD\,158926 \\
\hline
{\it Lyman doublets}\\
\multicolumn{2}{l}{Mg\,{\sc xii}\,$\lambda$8.419,8.425\AA } \\
$v$ (\kms)   & &&&& --222$\pm$200 \\
$FWHM$ (\kms)& &&&& 1046$\pm$475  \\
$F$          & &&&& 0.91$\pm$0.21 \\
He-to-H ratio& &&&& 12.5$\pm$3.1  \\
$T_{He/H}$ (keV)    & &&&& 0.42$\pm$0.02  \\ 
\multicolumn{12}{l}{Ne\,{\sc x}\,$\lambda$12.132,12.137\AA } \\
$v$ (\kms)   & &&& 1$\pm$314 & --72$\pm$43 & 1$\pm$147 & 4$\pm$192 && 283$\pm$80 & \\
$FWHM$ (\kms)& &&& 1111$\pm$836 & 1037$\pm$93 & 0 (fixed) & 0 (fixed) && 0 (fixed) & \\
$F$          & &&& 24.9$\pm$4.1 & 19.4$\pm$1.3 & 13.4$\pm$1.7 & 25.8$\pm$3.1 && 18.5$\pm$2.7 & \\
\multicolumn{12}{l}{O\,{\sc viii} Ly$\alpha$\,$\lambda$18.967,18.973\AA } \\
$v$ (\kms)   & --2$\pm$69 & 144$\pm$379 & 0$\pm$31 & --292$\pm$194 & --144$\pm$36 & 0$\pm$27 & --33$\pm$33 & 27$\pm$53 & 112$\pm$120 & --87$\pm$41 & --127$\pm$156\\
$FWHM$ (\kms)& 583$\pm$322 & 1750$\pm$1750 & 0 (fixed) & 0 (fixed) & 1267$\pm$69 & 0 (fixed) & 283$\pm$247 & 709$\pm$190 & 0 (fixed) & 0 (fixed) & 760$\pm$759 \\
$F$          & 57.1$\pm$4.2 & 12.3$\pm$2.2 & 126$\pm$8 & 42.2$\pm$5.4 & 261$\pm$11 & 97.0$\pm$3.2 & 159$\pm$5 & 109$\pm$5 & 22.4$\pm$3.2 & 218$\pm$13 & 32.4$\pm$4.2\\
He-to-H ratio   & 2.49$\pm$0.28 & 3.85$\pm$0.85 & 2.69$\pm$0.27 & 2.33$\pm$0.47 & 2.16$\pm$0.17         & 1.97$\pm$0.12         & 1.45$\pm$0.09         & 1.78$\pm$0.14   && 1.92$\pm$0.17        & 4.59$\pm$0.76 \\
$T_{He/H}$ (keV)        & 0.193$\pm$0.004 & 0.18$\pm$0.01 & 0.189$\pm$0.004 & 0.19$\pm$0.01 &       0.198$\pm$0.004 & 0.207$\pm$0.004       & 0.22$\pm$0.01 & 0.202$\pm$0.005 && 0.189$\pm$0.004      & 0.168$\pm$0.004\\
\multicolumn{12}{l}{O\,{\sc viii} Ly$\beta$\,$\lambda$16.006,16.007\AA } \\
$v$ (\kms)   & && --228$\pm$152 && --130$\pm$145 & 0$\pm$279 & 160$\pm$60 & 84$\pm$206 && --171$\pm$171 &\\
$FWHM$ (\kms)& && 0 (fixed) && 1296$\pm$159 & 1087$\pm$429 & 0 (fixed) & 933$\pm$602 && 0 (fixed) &\\
$F$          & && 14.7$\pm$3.6 && 34.0$\pm$3.3 & 11.3$\pm$1.8 & 26.0$\pm$3.4 & 14.4$\pm$2.5 && 34.0$\pm$4.5 &\\
\multicolumn{12}{l}{N\,{\sc vii}\,$\lambda$24.779,24.785\AA } \\
$v$ (\kms)   & --73$\pm$133 && --299$\pm$107 && 92$\pm$131 & 2$\pm$123 & 85$\pm$83 & --4$\pm$118 \\
$FWHM$ (\kms)& 439$\pm$495 && 0 (fixed) && 1838$\pm$233 & 0 (fixed) & 1264$\pm$254 & 729$\pm$535 \\
$F$          & 23.4$\pm$4.8 && 37.2$\pm$8.2 && 158$\pm$15 & 19.5$\pm$2.7 & 72.0$\pm$5.9 & 24.8$\pm$4.4 \\
He-to-H ratio   & &&&&&& 1.28$\pm$0.19 \\
$T_{He/H}$ (keV)        & &&&&&& 0.15$\pm$0.01 \\
\multicolumn{12}{l}{C\,{\sc vi}\,$\lambda$33.734,33.740\AA } \\
$v$ (\kms)   & && 64$\pm$86 &&& 0$\pm$60 & --122$\pm$49 &&&& 122$\pm$91 \\
$FWHM$ (\kms)& && 470$\pm$469 &&& 268$\pm$267 & 328$\pm$328 &&&& 0 (fixed) \\
$F$          & && 127$\pm$17 &&& 84.1$\pm$8.1 & 91.1$\pm$8.4 &&&& 79.9$\pm$13.6 \\
\hline
\multicolumn{12}{l}{Fe\,{\sc xvii}\,$\lambda$15.014,15.261\AA } \\
$v$ (\kms)   & && 1$\pm$50 && --115$\pm$36 & 9$\pm$49 & 0$\pm$56 & 188$\pm$130 & 0$\pm$65 & 104$\pm$166 & \\
$FWHM$ (\kms)& && 0 (fixed) && 1192$\pm$81 & 707$\pm$185 & 0 (fixed) & 927$\pm$477 & 0 (fixed) & 1385$\pm$482 & \\
$F$          & && 38.7$\pm$5.2 && 96.2$\pm$4.4 & 51.6$\pm$2.5 & 112$\pm$5 & 33.0$\pm$3.4 & 26.4$\pm$2.8 & 84.5$\pm$8.2 & \\
\multicolumn{12}{l}{Fe\,{\sc xvii}\,$\lambda$16.780,17.051,17.096\AA }\\
$v$ (\kms)   & && --373$\pm$154 && --585$\pm$42 & --474$\pm$58 & --602$\pm$26 & --490$\pm$86 & --614$\pm$147 & --665$\pm$56 & \\
$FWHM$ (\kms)& && 937$\pm$632 && 989$\pm$90 & 0 (fixed) & 0 (fixed) & 504$\pm$504 & 0 (fixed) & 0 (fixed) & \\
$F_{16.780}$  & && 17.1$\pm$3.9 && 40.2$\pm$4.2 & 18.8$\pm$1.9 & 40.3$\pm$3.5 & 8.28$\pm$2.07 & 5.42$\pm$2.03 & 21.6$\pm$4.9 & \\
$F_{both\,17}$& && 54.1$\pm$8.1 && 107$\pm$6 & 49.4$\pm$3.0 & 115$\pm$5 & 33.7$\pm$3.3 & 27.7$\pm$3.0 & 79.7$\pm$8.1 & \\
\hline
\end{tabular}
\\
\tablefoot{Fluxes are given in units 10$^{-6}$\,ph\,cm$^{-2}$\,s$^{-1}$, and correspond to fluxes for all components of Lyman doublets or of iron line groups. The provided ratios are uncorrected for interstellar absorption; since the $f$ flux of O\,{\sc vii} was set to zero for HD\,35468, no $R_{fir}$ was derived in this case. For the He-like triplet of Neon, the presence of multiple iron lines in the area may render the line fluxes unreliable because of blending with iron lines, hence they are provided for information only and the ratios are not indicated. Errors correspond to 1$\sigma$ uncertainties -- if asymmetric, the larger value is given here.}
}
\end{sidewaystable*}

We compared the results obtained for the different lines for each target, taking the errors\footnote{The errors on shifts in Table \ref{linefit} were derived from fitting errors and error propagation, hence {\it systematic} uncertainties are not included. While those linked to atomic parameters (ATOMDB) are difficult to estimate, it is known that, for example the RGS  wavelength scale displays variations with $\sigma\sim5$\,m\AA\ \citep{dev15}, corresponding to an additional error of 70--80\,km\,s$^{-1}$ on the positions of the oxygen lines.} into account but we detected no significant ($>3\sigma$) and coherent line shifts for any of them. There are systematic blueshifts for the iron lines near 17\AA\ for all targets, which remain unexplained and are probably not real. Excluding these blueshifts, we find that 59\% of measurements lie within $\pm1\sigma$ of zero, 83\% lie within $\pm2\sigma$, and 97\% lie within $\pm3\sigma$; these measurements match well the normal distribution centred on zero. Our targets {hence} display no line shifts. Similarly, the line widths appear overall negligible compared to the resolution for all targets except HD\,38771, for which $FWHM\sim1250$\,\kms, which is a value close to the wind terminal velocity \citep{cro06,sea08}. While other stars were observed using the RGS, which has a poorer spectral resolution than HEG/MEG, it should be noted that $FWHM$ larger than 1000\,\kms\ would have been detected, considering the errors and instrumental broadening. The X-ray lines thus generally appear quite narrow, confirming the results found for other B stars \citep[e.g.][]{wal07,coh08,fav09}.

We now turn to the line ratios: $(f+i)/r$ in He-like triplets and flux ratios between He-like and H-like lines are indicators of temperature while $\mathcal{R}=f/i$ is a probe of the formation radius of the X-ray emission. Correcting for the energy-dependent interstellar absorption is unnecessary for ratios involving the closely spaced $fir$ lines, but such a correction needs to be performed for the He-to-H-like flux ratios since the lines are more distant in this case. However, since our targets are nearby, their interstellar absorptions are small (Table \ref{journal}), hence these corrections are of very limited amplitude.

As in \citet{naz12}, we used the ATOMDB, although with the newer version 3.0.8, to compute the expected ratios He-to-H-like flux ratios and $(f+i)/r$ ratios as a function of temperatures, and compared them to the observed values (see results in Table \ref{linefit}). The derived temperatures, typically $\log(T)\sim6.35$, are in line with the lowest ones found in global fits (see next section) and are also comparable to temperatures observed in normal B stars \citep[e.g.][see also further below]{wal07,zhe07}. 

The negligible $f$ lines indicate that the emission from He-like triplets arises close to a source of UV, i.e. close to the stellar photosphere (e.g. \citealt{por01}). In fact, for massive stars, $\mathcal{R}=f/i=\mathcal{R}_{0}/(1+\phi/\phi_{\rm C})$. The $\mathcal{R}_{0}$ values were derived from ATOMDB calculations, and taken at the temperature indicated by the $(f+i)/r$ ratio (see above and Table \ref{linefit}). The $\phi_{\rm C}$ values were taken from \citet{blu72}. Finally, we used TLUSTY spectra \citep{lan03,lan07} in the UV domain with solar abundances to get the stellar fluxes in the velocity interval [--2000; 0] km s$^{-1}$ near the rest wavelengths of the 2$^{3}$S$_{1}$ $\rightarrow$ 2$^{3}$P$_{1,2}$ transition (for a similar approach, see \citealt{leu06}). We consider TLUSTY models with the closest temperatures to the ones listed in Table \ref{journal} and the closest surface gravities predicted by \citet{nie13} for the different spectral types of non-supergiant stars -- for HD\,38771, HD\,44743, and HD\,52089, the surface gravities derived by \citet{tou10} and \citet{fos15} were considered. A microturbulence velocity of $\xi$=10 km s$^{-1}$, which is a value adapted to our target types (e.g. \citealt{hun09}), was considered, unless models with the required temperatures and surface gravities were not available within the TLUSTY grid associated to $\xi$=10 km s$^{-1}$ (i.e. for HD\,35468, HD\,44743, HD\,52089, and HD\,158926). In this case, TLUSTY models with $\xi$=2 km s$^{-1}$ were taken into account. We performed tests to check the consistency of our choice: using these two microturbulence velocities resulted in similar formation radii, within error bars, for an effective temperature of 22000 K and a surface gravity of 3.00 dex. Since these UV fluxes are diluted for calculating the $\phi$ values, which can be represented by the multiplication by the dilution factor $W(r) = 0.5\,\{ 1 - [1 - (R_{\star}/r)^{2}]^{1/2} \}$, the $\mathcal{R}$ ratios constrain the position of the emitting plasma (see $R_{fir}$ in Table \ref{linefit}). Note that UV TLUSTY spectra only cover the 900 -- 3200 $\AA$ wavelength range, which therefore excludes the analysis of the Si\,{\sc xiii} triplet lines, hence $R_{fir}$ is not indicated in that case, but this concerned only HD\,38771 (and its $f/i$ ratio is very similar to that reported for HD\,37128, HD\,111123, or HD\,149438 by \citealt{wal07}). Focusing on the formation radii derived from O\,{\sc vii} lines (since they are available for all but one star), values ranging from $\sim$ 2 to $\sim$ 8 $R_{\star}$ are found. This is in good agreement with values determined for the same lines for the studied B stars mentioned in Sect. \ref{sect2}: for HD\,205021, \citet{fav09} derived a formation radius in the range $\simeq$ 3 -- 5 $R_{\star}$; for HD\,122451, \citet{raa05} estimated a formation radius $\lesssim$ 5.7 $R_{\star}$; \citet{naz08} gave a formation radius < 20 $R_{\star}$ for HD\,93030; \citet{mil07} derived a formation radius in the range 5 -- 8 $R_{\star}$ for HD\,116658; \citet{mew03} estimated the formation radius of HD\,149438 to be $\lesssim$ 10 $R_{\star}$; for HD\,37128, \citet{wal07} provided a formation radius of 10.2$\pm$2.6 $R_{\star}$.
\subsection{Global fits}

Having both low-resolution, broadband spectra (EPIC, ACIS-S 0th order) and high-resolution, narrower band spectra (RGS, HEG/MEG) allows us to investigate the full X-ray spectra in detail, deriving not only temperatures and absorptions, but also abundances. However, caution must be taken since abundances may vary wildly when all parameters are relaxed in one step. Therefore, we performed global fits in several steps. First, we fitted the low-resolution spectra with absorbed thermal emission models with fixed solar abundances \citep{asp09}. One thermal component was never sufficient to achieve a good fit, but two components were usually adequate, except for HD\,36960, HD\,79351, and HD\,144217, for which three components were needed to fit a high-energy tail. Fits with absorption in addition to the interstellar one were tried, but either the additional absorbing column reached a null value or the $\chi^2$ was worse than without such a column. This absence of (detectable) local absorption {is usual for B stars} \citep{naz11,rau15} and for the final fits we therefore considered only the interstellar contributions (see Table \ref{journal} for their values). Second, we fitted the high-resolution spectra with the same thermal models, but we fixed the temperatures while releasing abundances. {The abundance of an element is only allowed to vary if lines of that element are clearly observed} (and measured, see Table \ref{linefit}). In the third step, the temperatures were released again for fitting the high-resolution spectra. Finally, in the fourth step, a simultaneous fitting of both low-resolution and high-resolution spectra was performed with the same model as in the third step. The final results are provided in Table \ref{glofit}. The abundances found in the last two steps are similar, but they sometimes differ from those found in the second step, indicating that the formal error bars might actually be underestimated. 

\begin{sidewaystable*}
\centering
\caption{Results of the global fitting using models of the type $tbabs(ism)\times\sum vapec_i$. }
\label{glofit}
\setlength{\tabcolsep}{2pt}
\begin{tabular}{lccccccccccc}
\hline\hline
         & HD\,34816     & HD\,35468        & HD\,36512       & HD\,36960         & HD\,38771$^*$    & HD\,44743       & HD\,52089       & HD\,63922      & HD\,79351        & HD\,144217      & HD\,158926 \\
\hline
$kT_1$   & 0.186$\pm$0.004 & 0.172$\pm$0.004 & 0.116$\pm$0.008 & 0.081$\pm$0.001 & 0.215$\pm$0.010 & 0.197$\pm$0.001 & 0.199$\pm$0.002 & 0.190$\pm$0.002 & 0.223$\pm$0.016 & 0.188$\pm$0.003 & 0.174$\pm$0.006 \\
$norm_1$ & 8.63$\pm$0.42   & 2.91$\pm$0.10   & 19.8$\pm$3.7    & 13.6$\pm$0.78   & 46.5$\pm$4.3    & 7.66$\pm$0.28   & 9.93$\pm$0.40   & 13.3$\pm$0.5    & 4.07$\pm$0.23   & 27.6$\pm$1.21   & 5.37$\pm$0.40   \\
$kT_2$   & 0.41$\pm$0.07   & 0.59$\pm$0.03   & 0.291$\pm$0.004 & 0.459$\pm$0.015 & 0.52$\pm$0.02   & 0.558$\pm$0.006 & 0.579$\pm$0.004 & 0.488$\pm$0.017 & 0.74$\pm$0.02   & 0.481$\pm$0.019 & 0.48$\pm$0.06   \\
$norm_2$ & 0.46$\pm$0.30   & 0.19$\pm$0.02   & 7.00$\pm$0.55   & 1.20$\pm$0.04   & 7.30$\pm$0.82   & 1.91$\pm$0.05   & 5.78$\pm$0.13   & 1.78$\pm$0.13   & 3.73$\pm$0.09   & 4.18$\pm$0.35   & 0.43$\pm$0.09   \\
$kT_3$   &                 &                  &                 & 2.50$\pm$0.08 &                 &                 &                  &                & 1.60$\pm$0.03   & 4.4$\pm$6.2    &              \\
$norm_3$ &                 &                  &                 & 2.85$\pm$0.07 &                 &                 &                  &                & 3.73$\pm$0.09   & 0.32$\pm$0.10   &          \\
$\chi_{\nu}^2$ (dof) & 1.39(454)& 1.04(463)   & 1.31 (476)      & 1.35 (463)     & 1.08 (521)       & 1.54 (1414)     & 1.44 (1139)     & 1.21 (798)      & 1.40 (1020)     & 1.20 (377)      & 1.12 (312)      \\
$A_C$   &                               &                               & 0.61$\pm$0.16           &                               &                                & 1.69$\pm$0.16         & 1.26$\pm$0.13         &                                &                               &                                & 1.61$\pm$0.35 \\
$A_N$   & 1.07$\pm$0.13         &                               & 1.03$\pm$0.17         &                               & 1.05$\pm$0.33         & 1.27$\pm$0.09         & 2.28$\pm$0.15         & 1.21$\pm$0.11         &                               &                                 &    \\
$A_O$   & 0.54$\pm$0.03 & 0.45$\pm$0.03 & 0.97$\pm$0.11 & 2.74$\pm$0.15 & 0.33$\pm$0.03   & 0.78$\pm$0.03 & 0.76$\pm$0.03 & 0.69$\pm$0.03 & 0.34$\pm$0.03 & 0.78$\pm$0.04   & 0.74$\pm$0.07   \\
$A_N/A_C$ &                             &                               & 1.69$\pm$0.52           &                               &                               & 0.75$\pm$0.09           & 1.81$\pm$0.22         &                               &                                 &                               &    \\
$A_N/A_O$ &     1.98$\pm$0.26   &                               & 1.06$\pm$0.21         &                               & 3.18$\pm$1.04 & 1.62$\pm$0.13          & 3.00$\pm$0.23         & 1.75$\pm$0.18 &                               &                                 &    \\
$A_{Ne}$ &                & & 1.41$\pm$0.12 & 5.40$\pm$0.32 & 0.65$\pm$0.07 & 1.79$\pm$0.08 & 1.67$\pm$0.07 & 1.29$\pm$0.06 & 1.99$\pm$0.42 & 1.73$\pm$0.10 &  \\
$A_{Fe}$        &                               &                               & 0.83$\pm$0.06   &                               & 0.40$\pm$0.04         & 1.00$\pm$0.03   & 0.85$\pm$0.02         & 0.87$\pm$0.04         & 0.78$\pm$0.04         & 0.93$\pm$0.06         &  \\
$A_{O}/A_{Fe}$ &                        &                               & 1.17$\pm$0.16   &                               & 0.83$\pm$0.11         & 0.78$\pm$0.04   & 0.89$\pm$0.04 & 0.79$\pm$0.05         & 0.44$\pm$0.04 & 0.84$\pm$0.07   &  \\
$F^{\rm obs}_{\rm X}$           & 0.42$\pm$0.02         & 0.135$\pm$0.003         & 1.08$\pm$0.03         & 0.83$\pm$0.06         & 2.10$\pm$0.06         & 0.926$\pm$0.005         & 1.80$\pm$0.01         & 0.788$\pm$0.006       & 1.037$\pm$0.007         & 1.47$\pm$0.02         & 0.343$\pm$0.010 \\
{\tiny log($L_{\rm X}/L_{\rm BOL}$)} & $-7.24\pm0.02$ & $-8.48\pm0.01$ & $-6.77\pm0.01$ & $-6.76\pm0.03$  & $-7.26\pm0.01$ & $-7.597\pm0.002$ & $-7.421\pm0.002$ & $-7.044\pm0.003$ & $-6.767\pm0.003$ & $-7.38\pm0.01$ & $-8.21\pm0.01$ \\ 
{\tiny eq. {\it ROSAT} c.r.}     & 0.134 & 0.095 & 0.397 & 0.176 & 0.326 & 0.341 & 0.537 & 0.091 & 0.102 & 0.142 & 0.178  \\ 
\hline
\end{tabular}
\\
\tablefoot{Temperatures are in keV, normalisation factors in $10^{-4}$\,cm$^{-5}$, fluxes in the 0.5--10.0\,keV energy band in $10^{-12}$\,erg\,cm$^{-2}$\,s$^{-1}$, abundances $A_i$ in number, with respect to hydrogen, and with respect to solar values \citep{asp09}. Errors, derived from the Xspec {\it error} command for parameters and from {\it flux err} for fluxes, correspond to 1$\sigma$ uncertainties -- if asymmetric, the larger value is given here. EPIC spectra were fitted above 0.25\,keV and ACIS-S 0th order spectra above 0.35\,keV. \\$^*$For HD\,38771, there are three additional parameters: $A_{Mg}=0.44\pm0.06$, $A_{Si}=0.52\pm0.08$, and the Gaussian smoothing ({\it gsmooth}, convolved to the thermal model) with $\sigma=(1.23\pm0.05)\times 10^{-3}$\,keV at 6\,keV, corresponding to $FWHM=145\pm6$\,\kms. This smaller value, compared to individual line fits, may be explained by the grouping used for global fits, which blurs the line profiles.}
\end{sidewaystable*}

In general, the X-ray spectra appear very soft and the associated temperatures are low, i.e. 0.2--0.6\,keV, in line with results from \citet{raa05,zhe07,naz08,gor10} for optically bright, nearby B stars observed at high resolution in X-rays. However, in large surveys relying on low-resolution X-ray spectra, many B stars present higher temperatures \citep{naz09,naz11}, although the contamination by companions or magnetically confined winds cannot be excluded in such cases. In our sample, only three targets appear harder with a hotter component, i.e. HD\,36960, HD\,79351, and HD\,144217. The last two are binaries and we may suspect that this plays a role in their properties. {However, only longer observations including full monitoring} throughout the orbital period, would be able to reveal the exact origin of the hard X-rays{, which could be emitted by the companion or result from an interaction with it.} For HD\,36960, however, there is no obvious reason for such a hardness: it is not a known binary or magnetic object. Further investigation is thus needed to clarify this issue.

The X-ray luminosities of massive stars are generally compared to their bolometric {luminosities} (see penultimate line of Table \ref{glofit}). For single and non-magnetic O-type stars in clusters, the ratio $\log(L_{\rm X}/L_{\rm BOL})$ is close to $-7$ with a dispersion of about 0.2\,dex \citep[e.g.][]{naz11}. The situation appears more complex for early B stars. Surveys show a large scatter with ratios between --8 and --6 and the X-ray emission often appears at a constant flux level for the (rather few) detected cases \citep{ber96,naz11,naz14,rau15}; again, contamination of the X-ray emission by companions cannot be excluded. Furthermore, while magnetic O stars systematically appear brighter than their non-magnetic counterparts, this is not the case of magnetic B stars \citep[$\log{[L_{\rm X}/L_{\rm BOL}]}\sim-7.6$ for the faintest case in Fig. 4 and Table 7 of][]{naz14}, which further blurs the picture. However, our sample is rather clean in this respect, as the chosen B stars are nearby and not strongly magnetic (see Sect. \ref{sect2}). Our earliest B-type stars (B0--0.7) display ratios  between --6.8 and --7.4, in line with the O-star relationship. This confirms previous results for such objects \citep[and references therein]{rau15}. Four of the five latest B-type stars (B1--2.5) have ratios $<-7.4$, confirming the lower level of intrinsic X-ray emission of such (non-magnetic) stars found in {\it ROSAT} data \citep{coh97} and attributed to their weakest stellar winds. However, the last star, HD\,79351, appears overluminous. This can probably be explained by its binarity{; in this context, it is interesting to note that a flare} occurred during the observation of this star (see next section). 

In massive stars, X-rays are linked to stellar winds. The amount of X-ray emission can therefore be used to estimate the amount of mass loss heated to high temperatures. This is important information since, in the tenuous winds of B stars and late O stars, cooling times are long and hence the plasma remains hot once heated. Estimating mass-loss rates from optical/UV diagnostics may thus lead to underestimations. This was clearly demonstrated by \citet{hue12} for the late O star HD\,38666 (see also a similar discussion for HD\,111123 in \citealt{coh08}). We have used the best-fit normalisation factors (Table \ref{glofit}) to calculate the total emission measures $EM$s (since $norm=10^{-14}\,EM/[4\,\pi\,d^2]$), summing those of individual components (Table \ref{tableMdot}). These values were converted into rates of hot mass loss considering Eq. 1 of \citet{hue12} and its equivalent for constant velocity on p. 1868 of \citet[][correcting by the different stellar radii of our targets]{coh08}. This was carried out considering terminal wind velocities of 500, 1000, and 1500 km s$^{-1}$ and the formation radius derived for O\,{\sc vii} lines (Table \ref{linefit}). The derived range of values are provided in Table \ref{tableMdot}. We also computed mass-loss rates from the formulae of \citet[][listed in Table \ref{tableMdot}]{vin01}, using the temperature and bolometric luminosities of our targets listed in Table \ref{journal}, along with typical masses for the spectral types of our targets, taken from \citet{cox02}. Finally, mass-loss estimates derived from the analysis of optical spectra are also provided in Table \ref{tableMdot}, when available. Comparing all these values, we find that our targets fall into two categories. The hot plasma constitutes a small (or even negligible) part of the wind for HD\,36512, HD\,38771, HD\,44743, HD\,63922, and HD\,158926, or about half of the sample. For the other stars, the hot plasma appears as the dominant component of the circumstellar environment, as in HD\,38666 \citep{hue12}. In examining which stellar properties could justify this dichotomy, we find variable stars (see next section) and binaries in both categories; hence variability or multiplicity are not specific characteristics of stellar winds dominated by hot plasma. However, the three stars displaying a very hot component in their X-ray spectrum all belong to the second category, i.e. stars whose winds appear to be dominated by hot plasma. For such stars, the $EM$ may not perfectly reflect the intrinsic X-ray emission as there may be contamination by X-rays arising in a companion, in an interaction with it, or a still unknown phenomenon (see above). Further investigation is thus needed to ascertain that most of their wind truly is hot. In any case, our analysis suggests that some B stars have their winds in the X-ray emitting regime, such that using only optical/UV data could lead to underestimation of their actual mass-loss rate. We caution, however, that this result is preliminary, as detailed analysis of the optical/UV spectra are often not available for those stars.

\begin{sidewaystable*}
\centering
\caption{Derived mass loss from optical studies and estimates, from different prescriptions, of the hot mass loss derived with our X-ray data.}
\label{tableMdot}
\setlength{\tabcolsep}{2pt}
\begin{tabular}{lccccccccccccc}
\hline\hline
Target name                             & $\log(\dot{M}_{\rm literature})$      &\multicolumn{2}{c}{$\log(\dot{M}_{\rm Vink+01})$}&$EM$&$\log(\dot{M}_{\rm Huenemoerder+12})$  &$\log(\dot{M}_{\rm Cohen+08})$     \\
                                                & (M$_{\odot}$ yr$^{-1}$)       &\multicolumn{2}{c}{(M$_{\odot}$ yr$^{-1}$)}     &(10$^{52}$ cm$^{-3}$)  &(M$_{\odot}$ yr$^{-1}$)                &(M$_{\odot}$ yr$^{-1}$)              \\
                                                &                                       &       Low $T_{\rm eff}$ & High $T_{\rm eff}$      &                                       &                                               &                                               \\
\hline
HD\,34816       &                                       &                               & $-$8.50$\pm$0.06                &74.1$\pm$4.2           &<--8.1 to <--7.6               &<--8.0 to <--7.6                                       \\
HD\,35468       &                                       &$-$8.33$\pm$0.09       &                                       &2.20$\pm$0.07  &--10.0 to --8.4        $^a$            &--9.3 to --8.4 $^a$                            \\      
HD\,36512       &                                       &                               & $-$6.22$\pm$0.06                &2466$\pm$344   &<--7.2 to <--6.7               &<--7.1 to <--6.6       \\
HD\,36960        &                                      &                               & $-$7.79$\pm$0.07                &159$\pm$23             &<--7.8 to <--7.3               &<--7.8 to <--7.3       \\
HD\,38771       & $-$6.05       [1], $-5.92_{-0.07}^{+0.11}$ [2]                &$-$6.80$\pm$0.08       & $-$8.01$\pm$0.07        &252$\pm$21&--7.7 to --7.2      &--7.6 to --7.2 \\
HD\,44743       &                                       &$-$7.33$\pm$0.08       &$-$8.55$\pm$0.08               &26.1$\pm$0.8           &<--8.6 to <--8.2               &<--8.5 to <--8.0                               \\
HD\,52089       & $-9$ [3],$-8$ [4]             &$-$7.53$\pm$0.08       &                                       &28.9$\pm$0.8           &--8.3 to --7.8                        &--8.3 to --7.8 \\      
HD\,63922        &                                      &                               & $-$6.93$\pm$0.06                &460$\pm$16             &<--7.6 to <--7.1               &<--7.6 to <--7.1       \\
HD\,79351        &                                      &$-$8.71$\pm$0.10       &                                       &25.9$\pm$0.6           &--9.5 to --7.9        $^a$            &--8.7 to --7.8 $^a$                            \\
HD\,144217      &                                       &                               & $-$8.04$\pm$0.06                &59.1$\pm$2.3           &<--8.1 to <--7.7               &<--8.1 to <--7.6       \\
HD\,158926      &                                       &$-$6.50$\pm$0.08       &                                       &19.8$\pm$1.5           &--8.2 to --7.7                        &--8.1 to --7.7 \\
\hline
\end{tabular}
\\
\tablefoot{The second column lists the mass-loss rate value derived in recent optical studies using atmosphere models: [1]: \citet{cro06}, [2]: \citet{sea08}, [3]: Hamann et al. (priv. comm. with \citealt{fos15}), [4]: \citet{naj96}. The third and fourth columns provide the mass-loss estimates from the formulae of \citet{vin01} when the adopted $T_{\rm eff}$ is in the range 12500 -- 22500 K (cool side of the bi-stability jump) and 27500 $-$ 50000 K (hot side of the bi-stability jump), respectively. When the adopted $T_{\rm eff}$ (Table \ref{journal}) is in the range 22500 $-$ 27500 K (in the bi-stability jump), mass-loss estimates for both the cool and hot sides of the bi-stability jump are provided. The fifth column yields the total X-ray emission measures, estimated from the normalisation factors of the spectral fittings (Table \ref{glofit}). The last two columns provide the mass-loss rates predicted by Eq. 1 of \citet[][with $b$=0.97]{hue12} and the formula on p. 1868 of \citet{coh08}, respectively.  In these columns only a range of mass-loss rates are given: the lower values correspond to a terminal wind velocity of 500 km s$^{-1}$, while the highest values correspond to a terminal wind velocity of 1500 km s$^{-1}$. $^a$: These values were derived assuming $R_{fir}$ = 1 to 7 $R_{\star}$, which is in line with the formation radii derived in Sect. \ref{sec3_}.}
\end{sidewaystable*}

\begin{figure}
\center
\includegraphics[scale=0.87]{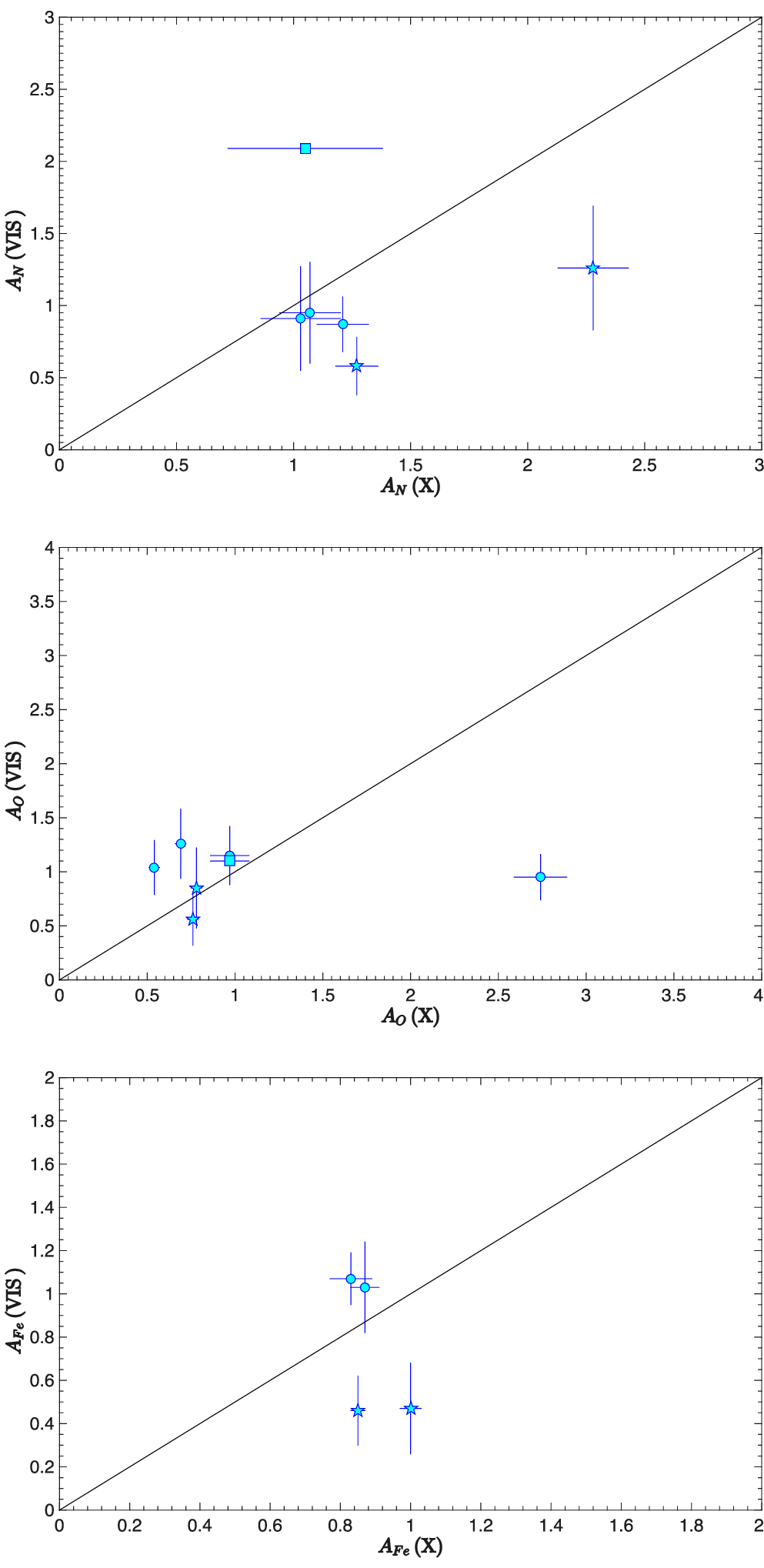}
\caption{Comparison between our N (upper panel), O (middle panel), and Fe (lower panel) abundance estimates and recent literature values obtained by optical studies. Circles, triangles, diamonds, squares, and star symbols represent values obtained by \citet{nie12}, \citet{ven02}, \citet{val04}, \citet{sea08}, and \citet{mor06,mor08}, respectively.}
\label{diffA}
\end{figure}

The abundances of our sample stars derived from our X-ray fits are listed in Table \ref{glofit} and those determined in optical studies from the literature are listed in Table \ref{optabund}. Figure \ref{diffA} provides a graphical comparison for the main elements (N, O, and Fe) for the stars in common. Except for the high O enrichment derived in X-rays for HD\,36960, which is not confirmed in the optical domain, the agreement between the derived abundances is fair (generally well within 3$\sigma$). Indeed, the formal fit errors are known to be smaller than the actual fit errors (e.g. \citealt{dep17}). An example of disagreement can also be found in the subsolar abundances found by \citet{coh08} and \citet{zhe07} for HD\,111123: even though they analysed the same dataset, the \citet{zhe07} abundances of O, Ne, and Mg are a factor of two smaller than those of \citet{coh08}; the origin of this discrepancy is unknown. We also note disagreements between optical studies, for example for HD\,38771 or HD\,52089, showing that improvements are required there too. In any case, not many comparisons between X-ray and optical abundances can be found in the literature. For HD\,205021, \citet{fav09} derived O, Si, and Fe abundances that are depleted compared to photospheric abundances. \citet{naz08} confirmed the depletion in C and O of HD\,93030 as well as its enrichment in N reported in the optical domain by \citet{hub08}. In fact, X-ray determinations would benefit from higher S/N data; for example, the X-ray abundances of HD\,66811, derived from very high-quality spectra, could be better constrained \citep{her13}.  

\begin{sidewaystable*}
{\tiny
\centering
\caption{Stellar abundances from optical studies. }
\label{optabund}
\setlength{\tabcolsep}{3.3pt}
\begin{tabular}{lccccccccccc}
\hline\hline
                        & HD\,34816                             & HD\,35468                                 & HD\,36512                             & HD\,36960                               & HD\,38771                                                     \\
\hline                          
{$A_C$}         & {0.68} [1], {0.89$\pm$0.15} [2]& {0.71} [3], {0.74} [1]               & {0.83$\pm$0.28} [2], {0.85$\pm$0.31} [7], {0.89$\pm$0.31} [6]   &{0.74$\pm$0.18} [9], {0.83$\pm$0.20} [2]                        & {0.11} [10], {0.20} [11]      \\                                                              
{$A_N$}         & {0.68} [1], {0.95$\pm$0.35} [2]& {1.82} [3], {2.00} [1]               & {0.89$\pm$0.31} [6], {0.91$\pm$0.36} [2], {1.00$\pm$0.26} [7]   &{0.76$\pm$0.16} [9], {0.78$\pm$0.22} [2]                        & {0.35$\pm$0.13}, 2.69$\pm$0.92 [10], {2.09} [11]       \\              
{$A_O$}         & {1.04$\pm$0.25} [2], {1.15} [1]& {}                                           & {1.05$\pm$0.34} [8], {1.06$\pm$0.41} [6], {1.15$\pm$0.27} [2]   &{0.95$\pm$0.21} [2], {1.07$\pm$0.21} [9]                        & {1.10} [11], {1.29$\pm$0.38} [10]            \\                                      
{$A_N/A_C$}    & {1.00} [1], {1.07$\pm$0.43} [2]& {2.56} [3], {2.70} [1]                & {1.10$\pm$0.57} [2], {1.00$\pm$0.49} [6], {1.18$\pm$0.53} [7]   &{0.94$\pm$0.35} [2], {1.03$\pm$0.33} [9]                        & {3.18$\pm$1.18};{24.5$\pm$8.4} [10], {10.45} [11]      \\              
{$A_N/A_O$}    & {0.59} [1], {0.91$\pm$0.40} [2]&                                               & {0.84$\pm$0.44} [6], {0.79$\pm$0.36} [2]                                        &{0.71$\pm$0.20} [9], {0.82$\pm$0.39} [2]                        & {0.27$\pm$0.13};{2.09$\pm$0.94} [10], {1.90} [11]               \\              
{$A_{Ne}$}      & {1.77$\pm$0.16} [2]           & {1.12$\pm$0.26}        [4]             & {1.52$\pm$0.43} [2], {1.70$\pm$0.55} [7]                                      & {1.59$\pm$0.54} [2]                                             & {}                                                                            \\                      
{$A_{Fe}$}      & {1.10$\pm$0.20} [2]           & {0.83}         [5]                            & {1.07$\pm$0.12} [2], {1.12$\pm$0.12} [7]                                        &{0.96$\pm$0.22} [2]                                             & {}                                                                            \\                              
{$A_{O}/A_{Fe}$}& {0.95$\pm$0.28} [2]           & {}                                            & {1.07$\pm$0.28} [2]                                                                     &{0.99$\pm$0.32} [2]                                             & {}                                                                            \\                      
{$A_{Mg}$}      & {}                                            & {}                                             & {}                                                                                                    & {}                                                                              & {0.40} [10]                                                             \\                      
{$A_{Si}$}      & {}                                            & {}                                             & {}                                                                                                    & {}                                                                              & {0.45} [10]                                                             \\                      
\hline\hline                     
                        & HD\,44743                             & HD\,52089                                                 & HD\,63922                                     \\
\hline
{$A_C$}         & {0.54$\pm$0.15} [12], {0.60} [1], {0.63} [3], {0.78$\pm$0.20} [14]            & {0.46$\pm$0.14} [12], {0.74$\pm$0.15} [14]                    & {0.74$\pm$0.15} [2]                     \\
{$A_N$}         & {0.58$\pm$0.20} [12], {0.76} [1], {0.79} [3], {0.83$\pm$0.24} [14]            & {0.81} [15], {1.26$\pm$0.43} [12], {2.15$\pm$0.42} [14]       & {0.87$\pm$0.19}  [2]                                    \\
{$A_O$}         & {0.85$\pm$0.37} [12], {1.15} [1], {1.23} [3], {1.10$\pm$0.28} [14]            & {0.56$\pm$0.24} [12], {1.02$\pm$0.30} [14]                    & {1.26$\pm$0.32}  [2]                                    \\
{$A_N/A_C$}    & {1.07$\pm$0.48} [12], {1.27} [1], {1.25} [3], {1.06$\pm$0.41} [14]            & {2.74$\pm$1.25} [12], {2.91$\pm$0.82} [14]                    & {1.18$\pm$0.35}     [2]                                                         \\
{$A_N/A_O$}    & {0.68$\pm$0.38} [12], {0.66} [1], {0.64} [3], {0.75$\pm$0.29} [14]            & {2.25$\pm$1.23} [12], {2.11$\pm$0.74} [14]                    & {0.69$\pm$0.23}     [2]                                 \\
{$A_{Ne}$}      & {0.91$\pm$0.34} [4]                                           & {0.85$\pm$0.24} [4]                                                     & {1.37$\pm$0.39}  [2]                                    \\
{$A_{Fe}$}      & {0.47$\pm$0.21}        [13]                                           & {0.46$\pm$0.16}  [12]                                                   & {1.03$\pm$0.21}     [2]                                 \\
{$A_{O}/A_{Fe}$}&                                                                               & {1.22$\pm$0.67}  [12]                                                   & {1.22$\pm$0.40}     [2]                         \\
\hline
\end{tabular}
\\
\tablefoot{Abundances are in number, with respect to hydrogen, and with respect to solar values, as for the X-ray {abundances} in Table \ref{glofit}. References: [1] \citet{gie92}, [2] \citet{nie12}, [3] \citet{ven02}, [4] \citet{mor08_1}, [5] \citet{val04}, [6] \citet{mar15}, [7] \citet{nie11}, [8] \citet{sim10}, [9] \citet{kil92}, [10] \citet{len91}, [11] \citet{sea08}, [12] \citet{mor08}, [13] \citet{mor06}, [14] \citet{fos15}, [15] \citet{fra10}.}
}
\end{sidewaystable*}

\section{Light curves}

\begin{figure*}
\includegraphics[width=6.26cm]{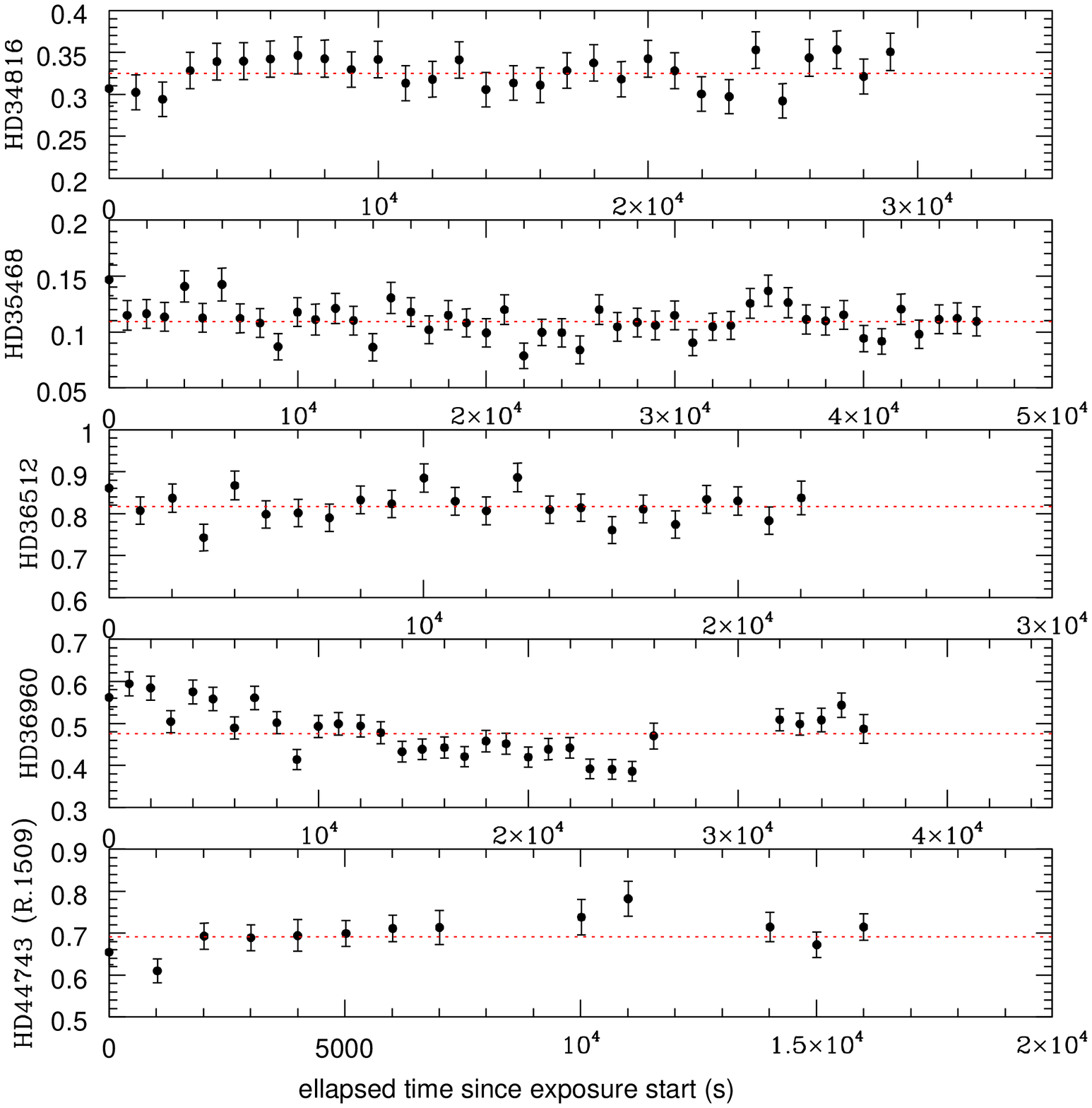}
\includegraphics[width=6.26cm]{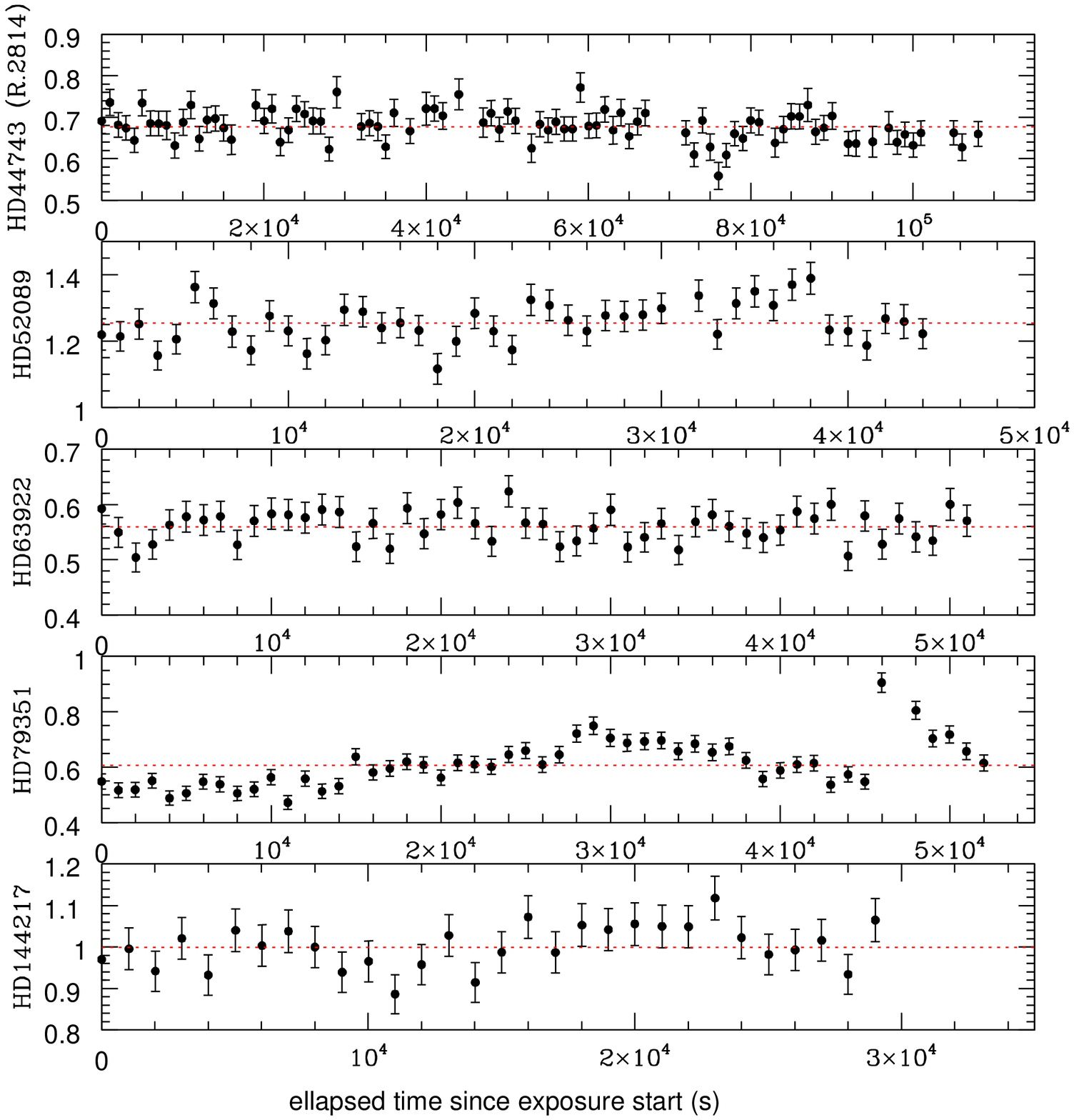}
\includegraphics[width=6.26cm]{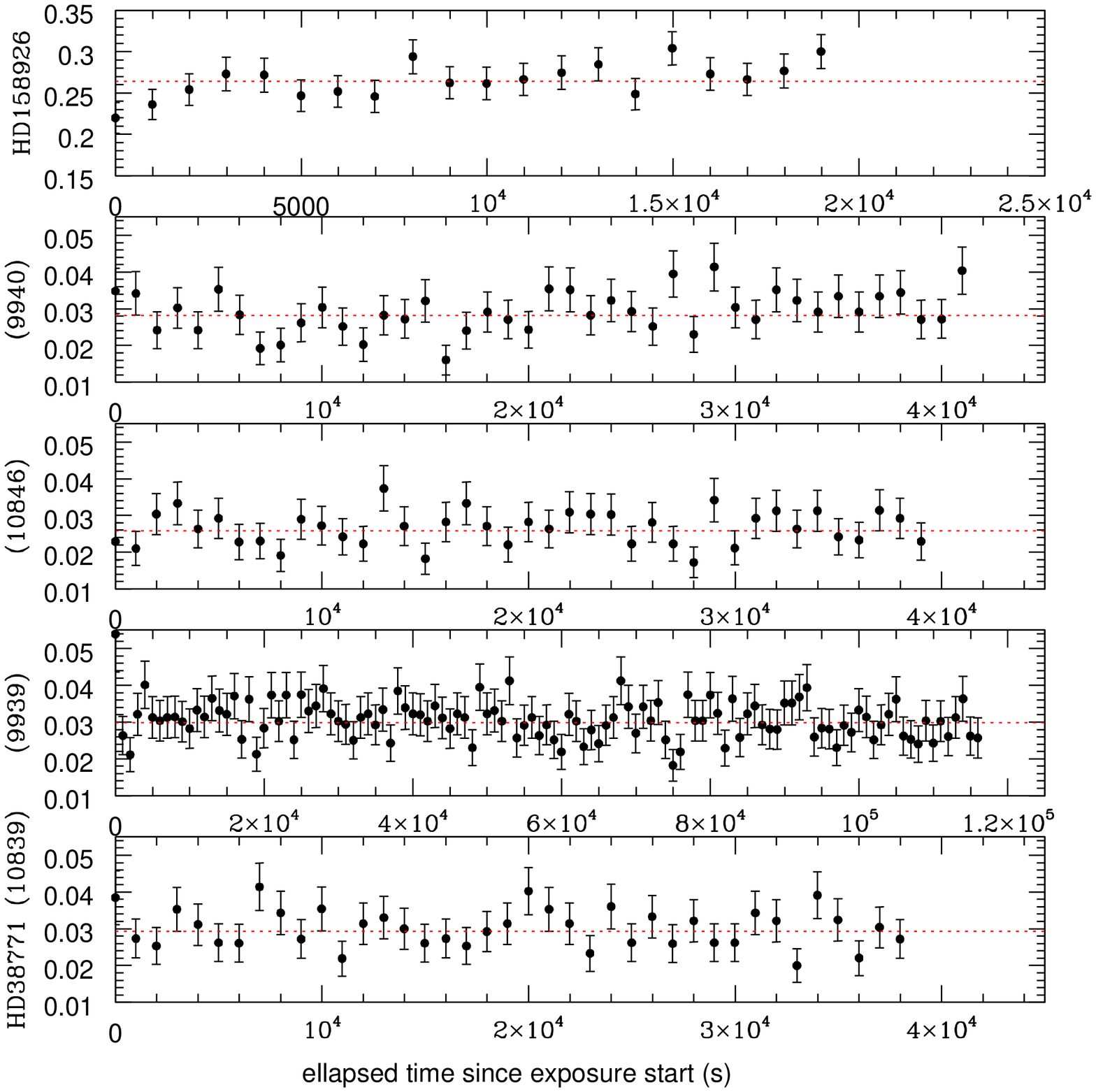}
\caption{Light curves in units cts\,s$^{-1}$ from low-resolution data (EPIC-pn, ACIS-S 0th order) for the different targets in the 0.3--10.0\,keV energy band and with 1ks bins. The weighted average is indicated with the dotted red line. }
\label{lc}
\end{figure*}

We finally examined the temporal evolution of the X-ray brightnesses of our targets.
The light curves in the 0.3--10.\,keV energy band are shown in Fig. \ref{lc}. We first performed $\chi^2$ tests for three different null hypotheses (constancy, linear variation, and quadratic variation). The improvement of the $\chi^2$ when increasing the number of parameters in the model (e.g. linear trend versus constancy) was also determined by means of Snedecor F tests \citep[nested models; see Sect.~12.2.5 in][]{lind76}. As threshold for significance, we used 1\% and we considered that a threshold of 10\% indicates only marginal evidence. This yields the following results. 

Two stars are clearly compatible with constancy, i.e. HD\,34816 and HD\,38771. For the latter, only the ACIS-S light curve with 1ks bins could be analysed, as there are not enough counts in each bin for a meaningful $\chi^2$ test of the 100s-binned light curves. We also analysed the light curves of each \ch\ exposure with the appropriate CIAO tool\footnote{http://cxc.harvard.edu/ciao/threads/variable/index.html\#glvary}, and the variability index was found to be zero (definitely not variable) for ObsID 9939, 10839, and 10846, and 2 (probably not variable) for ObsID 9940, which confirms the $\chi^2$ results. We also tested the combination of the four individual light curves using $\chi^2$ tests with the same null result as for individual light curves.

Four stars reveal marginal variability (i.e. presence of trends and/or rejection of constancy at the 1--10\% level), i.e. HD\,35468, HD\,44743, HD\,63922, and HD\,144217. HD\,35468 is detected to be variable for the MOS2 light curves, but this star is only marginally variable {in pn data} and appears to be compatible with constancy {in MOS1 data}. For HD\,44743, trends provide better fits for pn data in Rev. 1509 and for MOS1 and pn data for Rev. 2814. A marginal variability is further detected for all instruments in Rev. 2814, and when the two datasets are combined. The 100s-bin light curves of HD\,144217 are better fitted by trends. 

\begin{figure}
\center
\includegraphics[width=8.cm]{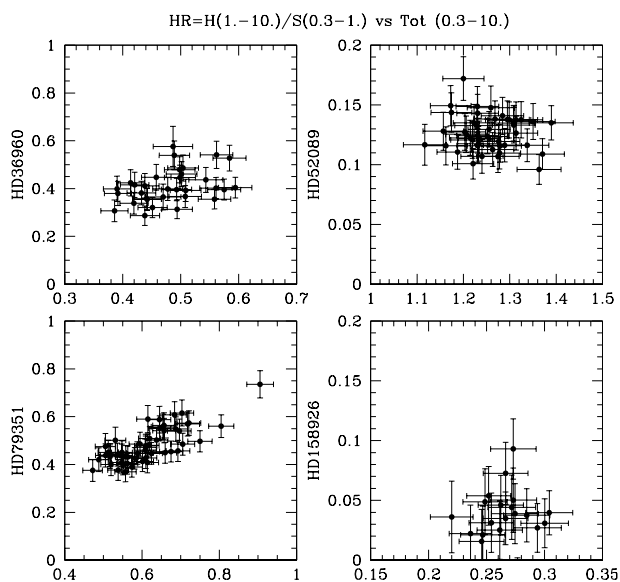}
\caption{Hardness ratios as a function of the count rates, in the 0.3--10.0\,keV energy band, and in units cts\,s$^{-1}$, for four clearly variable sources (for EPIC-pn and for 1ks bins). The hardness ratios are defined as the ratios between count rates in the 1.0--10.0\,keV and 0.3--1.0\,keV energy bands. The correlation coefficients amount to 40\% for HD\,36960, -9\% for HD\,52089, 81\% for HD\,79351, and 5\% for HD\,158926.}
\label{hr}
\end{figure}

Five stars exhibit a clear variability (significance level <1\%), i.e. HD\,36512, HD\,36960, HD\,52089, HD\,79351, and HD\,158926. HD\,36512 is detected to be variable for the 100s-bin pn light curve, and a trend provides a clearly better fit than a constant for MOS2 and pn. HD\,36960 appears significantly variable for all instruments, with trends -- especially parabolic trends -- always providing better fits. Indeed, the light curve presents a large oscillation with maxima at the beginning and end of the exposure and a minimum in between. This change in brightness is only marginally accompanied by a change in hardness (see Fig. \ref{hr}), however. HD\,52089 appears variable in pn data and a trend provides a clearly better fit than a constant. HD\,79351 appears significantly variable for all instruments. Indeed, the light curve steedily increases at first, then slightly decreases, and finally presents a flare-like activity. Furthermore, when it brightens, HD\,79351 also becomes harder (see Fig. \ref{hr}). HD\,158926 is detected to be significantly variable for all instruments when considering 100s bins, with trends providing a better fit to pn data.

\begin{figure*}
\center
\includegraphics[width=8.cm]{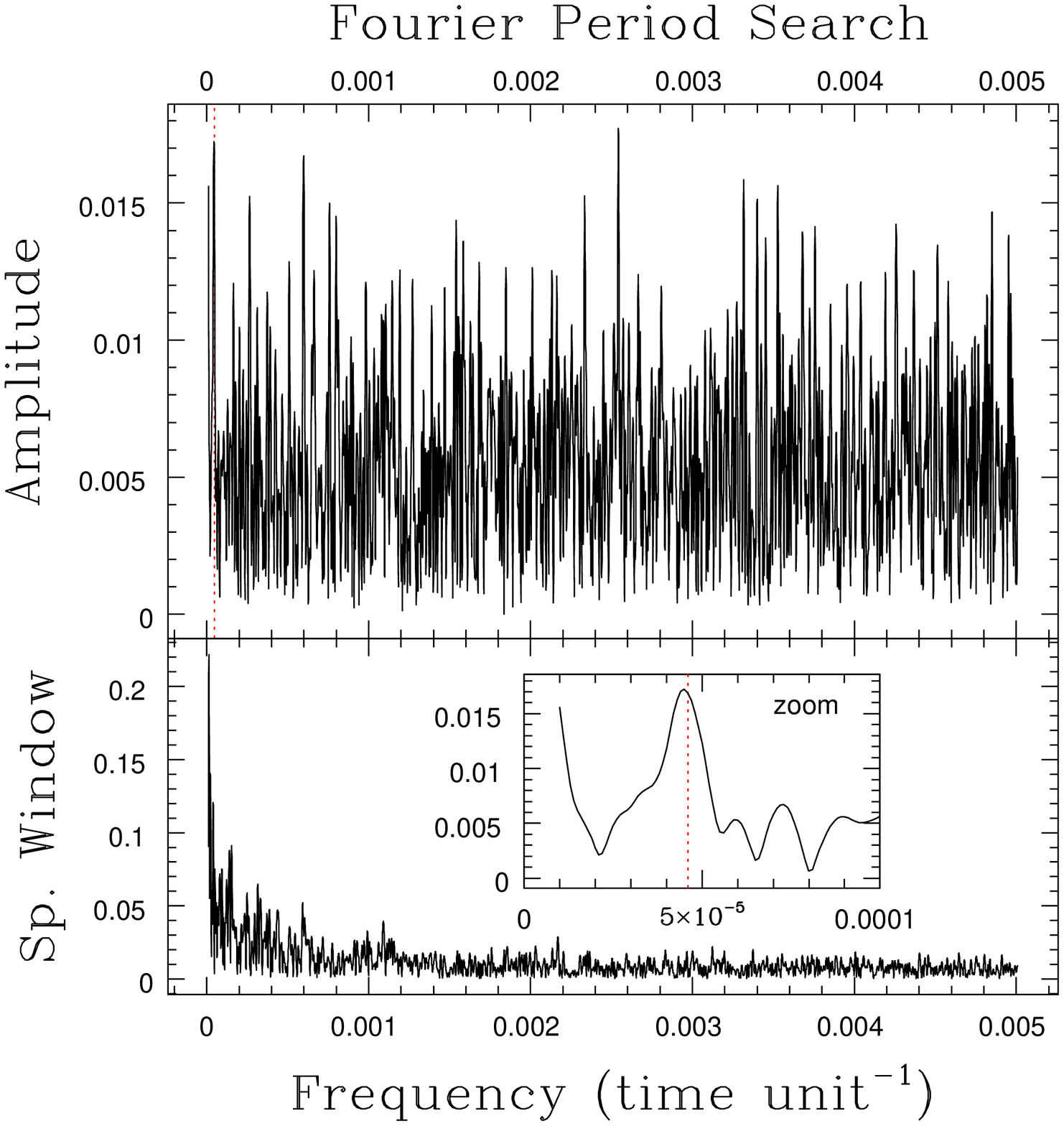}
\includegraphics[width=8.cm]{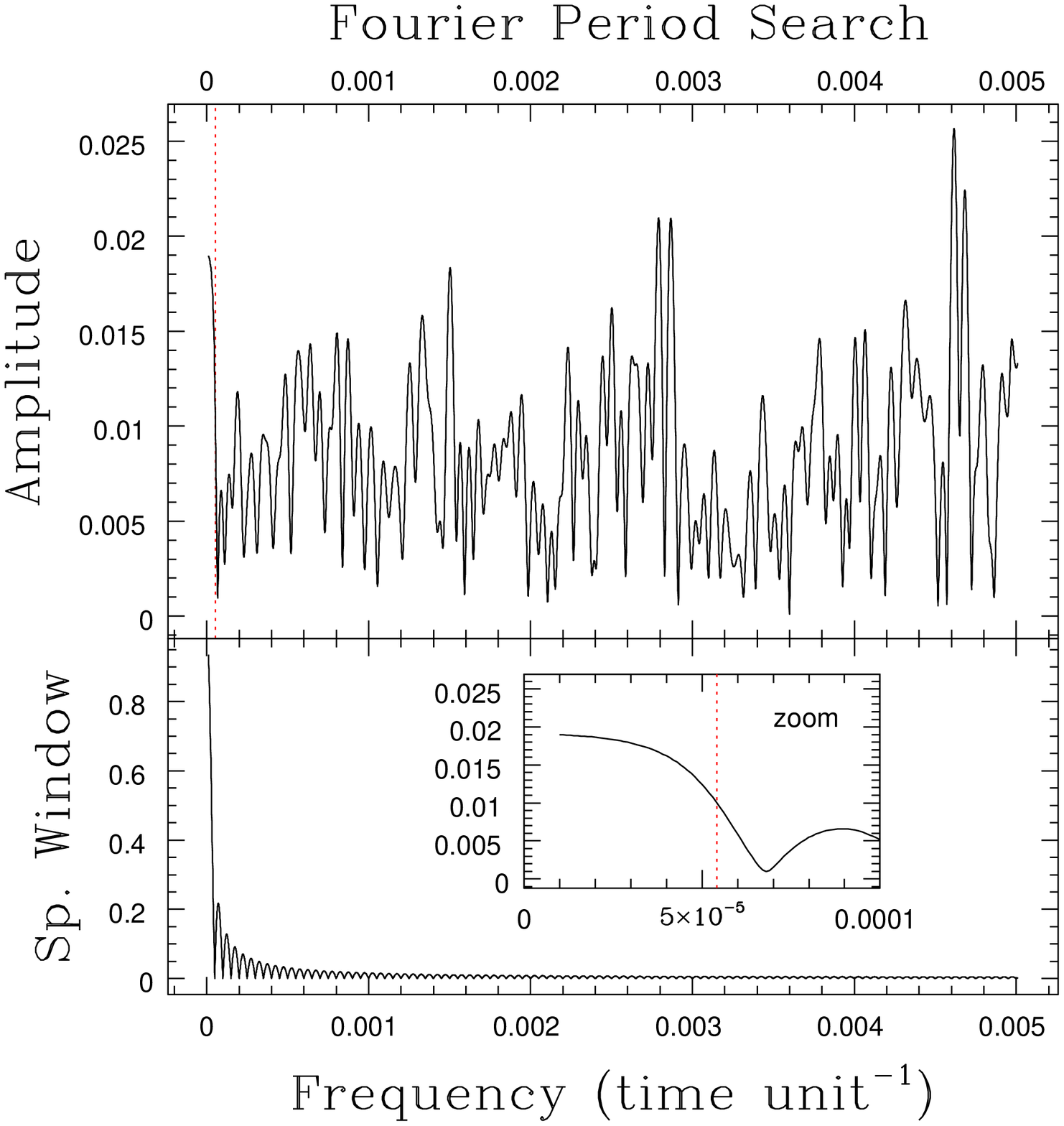}\\
\includegraphics[width=8.cm]{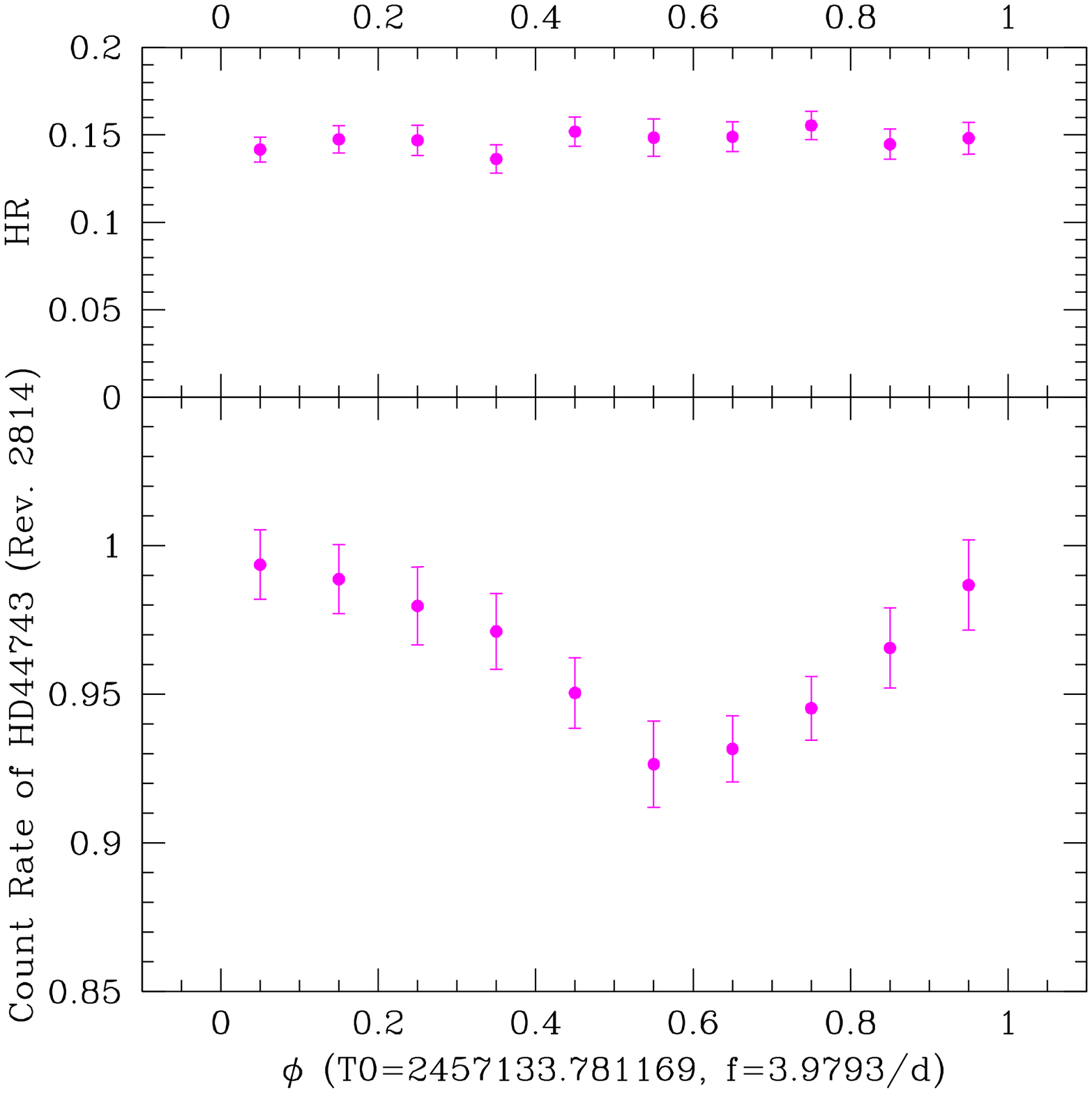}
\includegraphics[width=8.cm]{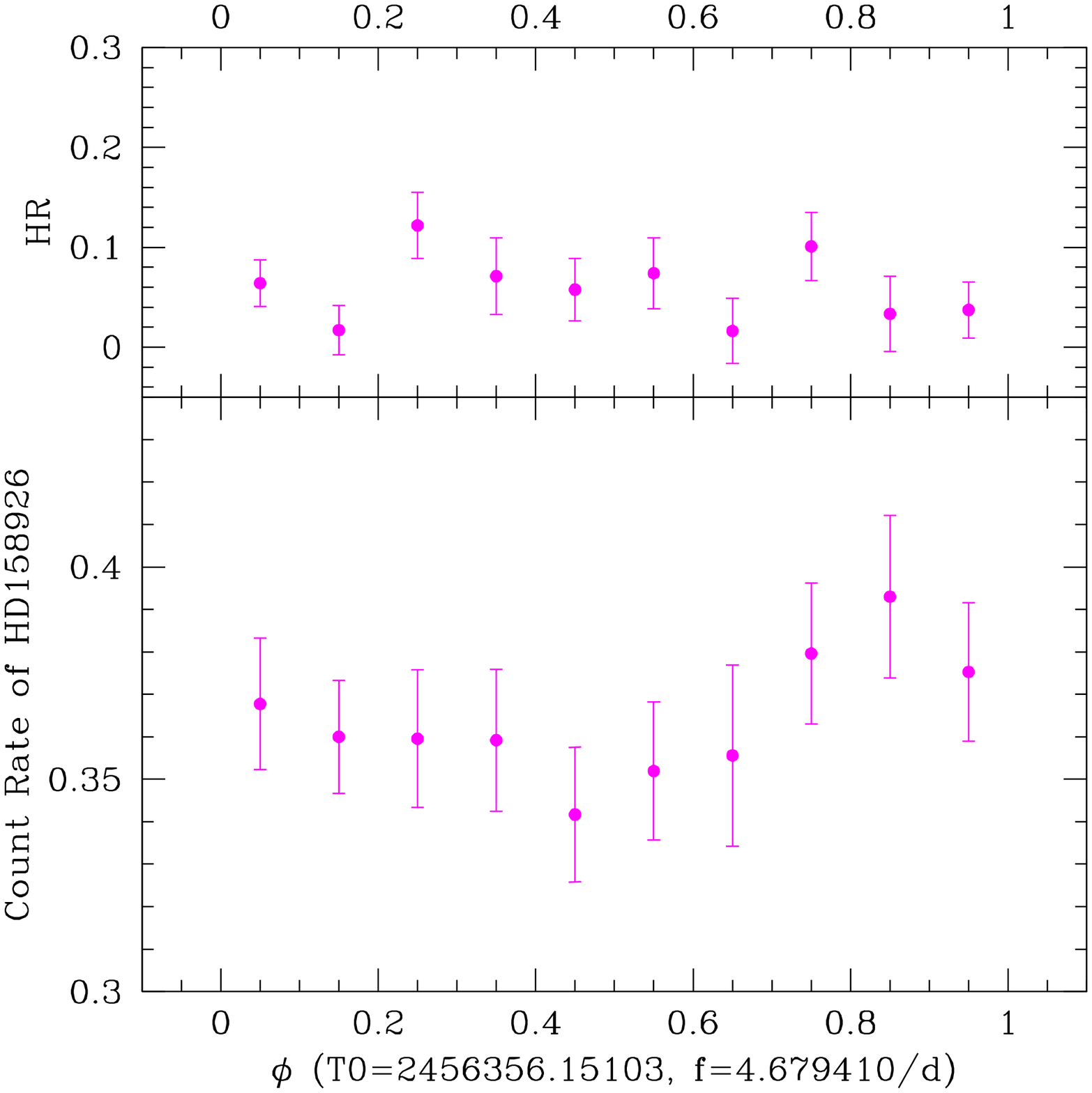}
\caption{{\it Top:} Periodograms based on the pn light curves with 100s bins of HD\,44743 (left, for Rev. 2814) and HD\,158926 (right), along with their associated spectral window. An inset provides a zoom on the lowest frequencies with the $\beta$\,Cephei pulsation frequencies shown by the red dotted lines.
{\it Bottom:} Light curves of HD\,44743 (left, for Rev. 2814) and HD158926 (right) folded using their dominant $\beta$\,Cephei frequency (see text).The values $T_0$ are arbitrary and were chosen to correspond to the start time of the \xmm\ observations (Table \ref{journal}). The count rates correspond to EPIC values (i.e. the addition of MOS1, MOS2, and pn count rates) while the hardness ratios, defined as in Fig. \ref{hr}, correspond to means (i.e. $[HR({\rm MOS1})+HR({\rm MOS2})+HR({\rm pn})]/3$). The phasing was here performed on light curves with 1ks bins; the 100s-binned light curves provide similar but slightly noisier results.}
\label{fig5}
\end{figure*}

Next, we applied period search algorithms \citep{hmm,gra13} to the light curves with 100s bins, for each EPIC camera, but no peak stands out both clearly and coherently (i.e. for all instruments) from the periodograms (although see further below). 

Two types of specific timescales exist for our targets: orbital periods and pulsation periods. The former are usually long compared to the exposure length (5.9\,d for HD\,158926, \citealt{ber00}; 6.745\,d for HD\,79351, \citealt{bus60}; 6.83\,d for HD144217, \citealt{hol97}), hence cannot be tested with the current dataset. However, we may still examine whether binarity may explain the flare-like behaviour of HD\,79351. Indeed, \citet{bus60} derived a mass function of 0.00663565 for this system. Assuming the mass taken from \citet[][$\sim 10\,M_{\odot}$]{cox02} for the primary star, we find a minimum mass of $\sim 0.7 M_{\odot}$ for the secondary star. It may thus be a PMS star, which could have undergone a flare during the observation. Indeed, the average X-ray luminosity of HD\,79351 is $\sim$3$\times$10$^{30}$ erg s$^{-1}$ (Table \ref{glofit}) and its brightness increased by less than a factor of 2 during the flare (Fig. \ref{lc}), which remains compatible with PMS flaring luminosities \citep{gud09}. However, a better knowledge of the HD\,79351 system is needed before its X-ray properties can be fully understood.

Regarding the latter timescales, it should be noted that the following two targets are known $\beta$\,Cephei: HD\,44743 \citep[three closely spaced frequencies with the strongest at $f=3.9793\pm0.0001$\,d$^{-1}$;][]{sho06} and HD\,158926 \citep[dominant frequency at $f=4.679410\pm0.000013$\,d$^{-1}$;][]{uyt04b}. For HD\,44743, a peak exists close to that dominant timescale in the periodogram of EPIC data taken in Rev. 2814 (top of Fig. \ref{fig5}). The amplitude of {this peak} is not very large but it is present and in fact, if the background flares are not discarded hence longer and more complete light curves are available, {this peak} stands out very clearly. In this context, it should be remembered that the known pulsators HD\,122451 ($\beta$\,Cen), HD\,116658 (Spica), HD\,205021 ($\beta$\,Cep), and HD\,160578 ($\kappa$\,Sco) do not display X-ray variations linked to their pulsations \citep{raa05,mil07,fav09,osk15}, but HD\,46328 ($\xi^1$\,CMa) does \citep{osk14,naz15} and $\beta$\,Cru may \citep[though see refutation in \citealt{osk15}]{coh08}. Therefore, even if periodograms do not show very strong peaks at these periods, we performed a folding using these known frequencies as a last check (bottom of Fig. \ref{fig5}). For HD\,44743, the presence of a modulation is clearly confirmed with a peak-to-peak amplitude of 14\% (corresponding to five times the error on individual bins); {this modulation} is not accompanied by significant hardness changes, but neither did HD\,46328. The case of HD\,158926 appears less convincing as the changes are not coherent from one instrument to the other, and the combination of all three EPIC instruments only leads to a slight modulation (the peak-to-peak variation is only 3$\sigma$), clearly calling for confirmation before detection can be claimed. 

As a final exercise, we folded our best-fit spectral models (Table \ref{glofit}) through the {\it ROSAT} response matrices and derived the equivalent {\it ROSAT} count rates of our targets (reported in the last line of Table \ref{glofit}). Comparing our results to values tabulated by \citet{ber96}, we find negligible ($<3\sigma$) differences for all but three targets; HD\,35468 and HD\,158926 has become fainter by a factor of two, while HD\,36512 has brightened by 50\%. These three stars thus appear variable on both long and short timescales. While this could probably be linked to binarity for HD\,158926, there is no clear explanation for the other two. A monitoring of all three stars will then be needed to better understand their behaviour, in particular searching for a putative periodicity.

\section{Discussion and conclusions}
In this paper, we analysed the X-ray data of 11 early B stars. Combined with 8 other B stars previously analysed, our results provide the first X-ray luminosity-limited survey at high resolution, thereby constituting a legacy project. 

In this work, we performed line-by-line fitting, global spectral fitting (with temperatures, abundances, and brightnesses as free parameters), and variability analyses.

In many ways, our results confirm previous studies. B stars typically display soft and moderately intense X-ray emissions, their X-ray lines appear rather narrow and unshifted, and the X-ray emission arises at a few stellar radii from the photospheres.

The abundances we derived are in fair agreement with those found in optical data, taking errors into account. The X-ray brightnesses could be used to evaluate the total quantity of hot material surrounding the stars. Compared to expected mass-loss rates or values derived from the analysis of optical data, we find that in half of the cases, the hot mass-loss constitutes only a small fraction of the wind.

Roughly a quarter of our sample (3/11) display a high-energy tail and roughly half of our targets (6/11) display significant variations on short or long timescales. These properties are not mutually exclusive as two targets (HD\,36960 and HD\,79351) combine both specificities. Three out of the four binaries present at least one of these peculiarities, which thus appear not restricted to, but more common in, binaries. 

Finally, we also analysed in detail the temporal behaviour of two $\beta$\,Cephei pulsators. HD\,158926 presents flux changes on both short and long timescales, but they cannot be undoubtly assigned to the pulsational activity. On the contrary, HD\,44743 appears marginally variable in $\chi^2$ tests, but folding its light curve clearly reveals coherent variations with the optical period. This makes the star the second secure case of X-ray pulsator after HD 46328.

{There remains} much to be done, such as longer monitoring of the variable sources and more precise estimates of the parameters (formation radii and abundances). In this context, the advent of the Advanced Telescope for High-ENergy Astrophysics/X-ray Integral Field Unit (ATHENA/X-IFU) will certainly provide higher quality high-resolution spectra of B stars, which will further advance our understanding of their X-ray properties.
 
\begin{acknowledgements}
We thank Thierry Morel and Gregor Rauw for their useful comments.

We acknowledge support from the Fonds National de la Recherche Scientifique (Belgium), the Communaut\'e Fran\c caise de Belgique, the PRODEX \xmm\ contract, and an ARC grant for concerted research actions financed by the French community of Belgium (Wallonia-Brussels Federation). ADS and CDS were used in preparing this document. 
\end{acknowledgements}

\end{document}